\documentclass[%
 aip,
 amsmath,amssymb,
 reprint,%
]{revtex4-1}

\usepackage{graphicx}
\usepackage{dcolumn}
\usepackage{bm}

\usepackage[utf8]{inputenc}
\usepackage[T1]{fontenc}
\usepackage{mathptmx}

\begin{document}


\title{Beyond Graphene: Low-Symmetry and Anisotropic 2D Materials}

\author{Salvador Barraza-Lopez}
\affiliation{Department of Physics, University of Arkansas, Fayetteville, AR 72701, USA}

\author{Fengnian Xia}
\affiliation{School of Engineering and Applied Science, Yale University, New Haven, CT 06520, USA}

\author{Wenjuan Zhu}
\affiliation{Department of Electrical and Computer Engineering, University of Illinois, Urbana, IL 61801, USA}

\author{Han Wang}
\affiliation{Department of Electrical and Computer Engineering, University of Southern California,  Los Angeles, CA 90089, USA}

\date{\today}

\maketitle

Low-symmetry 2D materials---such as ReS$_2$ and ReSe$_2$ monolayers, black phosphorus monolayers, group-IV monochalcogenide monolayers, borophene, among others---have more complex atomistic structures than the honeycomb lattices of graphene, hexagonal boron nitride, and transition metal dichalcogenides. The reduced symmetries of these emerging materials give rise to inhomogeneous electron, optical, valley, and spin responses, as well as entirely new properties such as ferroelasticity, ferroelectricity, magnetism, spin-wave phenomena, large nonlinear optical properties, photogalvanic effects, and superconductivity. Novel electronic topological properties, nonlinear elastic properties, and structural phase transformations can also take place due to low symmetry. The ``Beyond Graphene: Low-Symmetry and Anisotropic 2D Materials'' Special Topic was assembled to highlight recent experimental and theoretical research on these emerging materials.

\section{Introduction}\label{sec:intro}
Graphene,\cite{RevModPhys.81.109,RMP2,RMP3} hexagonal boron nitride monolayers,\cite{hBN,hBN2} and transition-metal dichalcogenide monolayers (TMDCs) with a 2H symmetry (2H-TMDCs)\cite{mak1} are all well-established two-dimensional (2D) materials. Graphene displays a sixfold rotational symmetry and also has three mirror planes, while hBN and hexagonal 2H-TMDCs have a sixfold roto-inversion symmetry with two mirror planes. The physical properties of these materials have been studied at length (see, {\em e.g.,} References \onlinecite{Katsnelson}, \onlinecite{avouris}, and \onlinecite{TMDCs}). Along these lines, and appearing in the ``Beyond Graphene: Low-Symmetry and Anisotropic 2D Materials'' Special Topic, {\em Yamasue and Cho} use scanning nonlinear dielectric microscopy to visualize unintentional carrier doping of few-layer Nb-doped MoS$_2$.\cite{doping} They show that atomically thin layers exhibit a p- to n-type transition as the thickness decreases.  This sensitive technique is applicable to arbitrary two-dimensional materials, and it will advance understanding of and the ability to predict device characteristics even at an early stage of the fabrication process.

But two-dimensional structural anisotropy implies that a given material displays different physical properties when probed along different spatial directions, and lowering the symmetry of graphene and of other two-dimensional materials by the application of strain leads to remarkable effects, not available in the highly-symmetric phase. For example, graphene develops local gauge fields whereby electrons behave as if under an external magnetic field,\cite{VOZMEDIANO2010109,maria,Review} and its spin-orbit coupling strength can also be tuned by curvature.\cite{PhysRevB.74.155426} Uniaxial strain in turn induces a piezoelectric response in binary materials such as hexagonal boron nitride\cite{duerloo,piezohBN} and 2H-TMDCs.\cite{duerloo,piezoMoS2} Local strain has also been proposed to tune the electronic band gap in 2D semiconductors.\cite{funnel} Another striking effect from a lowered local symmetry is the superconductivity observed in angularly-mismatched (moir\'e) graphene bilayers.\cite{twisted}

While previous examples point to an engineered anisotropy, new 2D materials with a spontaneous, {\em intrinsic} lower symmetry are being predicted and/or experimentally discovered.\cite{C2DB,MarzariDatabase,Hennig2017,scirep2017,duerloo2D,2DMatPedia} Their structural anisotropy influences all possible (electric, magnetic, optical, or mechanical) material responses, and the Special Topic ``Beyond Graphene: Low-Symmetry and Anisotropic 2D Materials'' in the {\em Journal of Applied Physics} has been assembled to showcase recent research into a large number of 2D materials displaying an intrinsic structural anisotropy. As shown by the breadth of submissions, anisotropic 2D materials represent an exciting and extremely active avenue of research in Physics, Chemistry, and Engineering.

\section{Layered and 2D materials with an in-plane intrinsic structural anisotropy}
\subsection{Transition metal dichalcogenides in the T and T' phases}
Metal atoms sit at prismatic positions In 2H-TMDCs, but they occupy the octahedral positions between two chalcogen layers in the trigonal (T) phase. Representing an early demonstration of permanent structural phase transitions onto a crystalline structure with a lower symmetry, a triclinic (T) phase was created at the surface of TMDC tantalum diselenide back in the mid-nineties.\cite{tmd0,tmdm1} TMDCs with trigonal symmetry have been shown to host superconductivity,\cite{tp_super} charge density waves,\cite{tp_cdw} quantum spin Hall semimetal behavior,\cite{qian_science_2014,fei_np_2017,tang_np_2017,wu_science_2018,wte2} and ferroelectricity.\cite{yuan_nc_2019,wte2}

Reporting their results in the ``Beyond Graphene'' Special Topic, {\em Saha et al.} undertook a systematic study of pressure induced lattice expansion and phonon
softening in layered ReS$_2$, a TMDC with trigonal structure. The techniques employed include $x-$ray diffraction and Raman spectroscopy. They observed all the eighteen active Raman modes in their experimental results under standard temperature and pressure conditions, and ascertained an iso-structural transition onto the 1T' phase (which is still trigonal but features an even more reduced symmetry) taking place above 6.1 GPa. The 1T' phase remained stable up to a pressure of 42 GPa. The softening of Raman modes was assigned to vibrational modes predominantly created by rhenium atoms.\cite{saha}

In addition, the scanning tunneling microscopy/spectroscopy work by {\em Plumadore and coworkers} showcases the properties of a graphene/ReS$_2$ heterojunction, in which novel properties become enabled by a combination of proximity effects and moir\'e patterns.\cite{plumadore} They observe a striped superpattern created by interlayer interactions between graphene’s hexagonal structure and the triclinic, low in-plane symmetry of ReS$_2$. They compared their experimental results with a theoretical model that estimates the
shape and angle dependence of the moir\'e pattern between graphene and ReS$_2$. Their results shed light on the complex interface phenomena between van der Waals materials with different lattice symmetries.

{\em \L. Kipczak and collaborators} studied the photoluminescence (PL) and Raman scattering properties of ultrathin ReSe$_2$---whose thicknesses ranged from nine to one monolayer---at 5 K and at room temperature,\cite{Kipczak} paving the way for the identification of few-layer ReSe$_2$ samples by optical means. The PL spectra of ReSe$_2$ layers display two well-resolved emission lines, which blue-shift by about 120 meV when the layer thickness decreases from nine monolayers to a monolayer, confirming a direct optical transition. More specifically, the two phonon modes of intralayer vibrations observed in Raman scattering spectra at about 120 cm$^{-1}$ exhibit an opposite evolution as a function of layer thickness. Their energy difference can serve as a convenient and reliable tool to determine the thickness of ReSe$_2$ flakes in the few-layer limit.\cite{Kipczak}

\subsection{Ultrathin black phosphorus (including monolayers)}

Black phosphorus (BP) is another layered material with in-plane anisotropy.\cite{avouris,ph1,ph2,ph3,Ling4523} Phosphorus belongs to the nitrogen group, which is located to the right of the carbon group in the Periodic Table of the Elements. Phosphorus has five valence electrons, while carbon has four. As it turns out, a black phosphorus monolayer and graphene are both three-fold coordinated, which means that they form three strong chemical bonds. In the case of graphene, the remaining ($\pi$) electron hovers out of plane. But a {\em lone pair} ensues in black phosphorus, leading to an out-of-plane buckling of the atoms in its unit cell, which turns {\em rectangular} and contains four atoms. This difference in the chemistry of carbon and phosphorus leads to a large number of anisotropic properties (elastic, electronic, and optical) observed on this material.\cite{avouris,2019re,xia2}

Publishing their results in the ``Beyond Graphene'' Special Topic, {\em Doha et al.} created an anti-reflection cavity that optimizes absorption in a BP layer, which was characterized using scanning photocurrent microscopy. They also modeled the devices by solving Maxwell’s equations and the drift–diffusion equation to obtain the optical absorption and photocurrent density in response to pulsed laser excitation. They observed a strong absorption of 36\% at 780 nm, which suggests a promising outlook for the THz performance of these devices.\cite{DohaBPAntenna} Additionally, computational work by {\em Sibari and collaborators} explores the relation among atomistic structure and electronic properties of few-layer black phosphorene,\cite{sibari} while {\em Betancur-Ocampo and coworkers} employed a Green's function formalism on a tight-binding model of a black phosphorus monolayer pnp junction, as well as a continuum description, and determine that these junctions operate as electron waveguides.\cite{betancur}

\subsection{Multiferroic behavior in layered and 2D materials with low symmetry}

Layered ferroelectrics---such as In$_2$Se$_3$,\cite{poh_nl_2018_in2se3} CuInP$_2$S$_6$,\cite{cips} BA$_2$PbCl$_4$,\cite{ba2pbcl4} 1T'-MoTe$_2$,\cite{yuan_nc_2019} and 1T'-WTe$_2$\cite{wte2}---provide unprecedented freedom for the design and fabrication of functional (van der Waals) heterostructures. This Special Issue features four works in a subset of these materials; namely, ferroelectric and ferroelastic group-IV transition monochalcogenide monolayers (MX, with M being Ge, Sn, or Pb, while X could be S, Se, or Te).\cite{2019re,colloquium} MX monolayers are a family of novel two-dimensional (2D) materials whose atomistic structures closely resemble the black phosphorus lattice. Most MX monolayers exhibit a broken inversion symmetry and are ferroelectric with a reversible in-plane electric polarization. MX monolayers are promising materials for applications in non-linear optics, photovoltaics, spintronics, and valleytronics.

The ``Beyond Graphene'' Special Topic features a Perspective Article by {\em Chang and Parkin}, in which a detailed exposition of the experimental creation and characterization of MX monolayers is provided.\cite{chang} Due to their relatively large exfoliation energy, the creation of MX monolayers is not an easy endeavor, which hinders the integration of these materials into the fast-developing field of 2D material heterostructures. They review recent developments in experimental routes to the creation of these materials, including molecular beam epitaxy and two-step etching methods. Other approaches that could be used to prepare MX monolayers, such as liquid phase exfoliation and solution-phase synthesis, were discussed as well. Quantitative comparisons between the material properties observed were also presented.\cite{chang}

In turn, {\em Gomes and Carvalho} provide a Tutorial of the electronic and optical properties of 2D group-IV monochalcogenides, including predictions from first-principles DFT calculations, and available experimental observations. They discuss the variation of the bandgap from bulk down to monolayer, and the respective band structures, which are characterized by multiple valence and conduction band valleys, making these materials suitable for a variety of applications, including valleytronics. They also discuss the emergence of spin-orbit splitting, piezoelectricity, and ferroelectricity as a result of the polar character of the monolayers. Current predictions of carrier mobilities in monolayers, and their potential application as thermoelectric materials were discussed as well. Ferroelectric and ferroelastic materials have thermally accessible elastic energy barriers separating degenerate structural ground states. {\em Du and collaborators} studied the effect of charge doping on the elastic energy barrier created by a Pnm2$_1 \to$ P4/nmm two-dimensional structural transformation of a black phosphorus monolayer, and nine ferroelectric/ferroelastic group-IV monochalcogenide monolayers. Group-IV monochalcogenide monolayers show a tunable elastic energy barrier for small amounts of doping: a decrease (increase) of the energy barrier can be engineered under a modest hole (electron) doping of no more than one tenth of an electron or a hole per atom. These results provide further guidance concerning a possible tunability of the ferroelectric-to-paraelectric transition temperature of these compounds by charge doping.\cite{doping2} Lastly, {\em Seixas} employed first-principles techniques to study the structural, electronic, and vibrational properties of fifteen group-IV monochalcogenide monolayers based on Janus substitution. These Janus materials are potential candidates for similar applications but with additionally broken symmetry that can enrich their electronic and optical properties.\cite{seixas}

\subsection{Borophene}
In the ``Beyond Graphene'' Special Topic,  {\em Sandoval-Santana and collaborators} address the dynamics of charge carriers in borophene with an Pmmn symmetry obeying an anisotropic Dirac Hamiltonian, subjected to illumination by linearly polarized light of arbitrary intensity. To this end, they develop analytical methods, including a set of unitary transformations that enable the reduction of the matrix-differential equation into a scalar differential equation, the Floquet theorem, and a Fourier spectral decomposition. They show that the quasi-energy spectrum develops an anisotropic structure in the intense field regime.\cite{gerardo}

\section{Additional anisotropic layered and 2D materials in the horizon}
Displaying the sheer vitality of this research field, {\em Luca Vannucci and coworkers}\cite{1-Vannucci} report a high-throughput search of anisotropic two-dimensional materials from the C2DB database,\cite{C2DB} which contains in excess of 1,000 entries. They give special attention to the ternary orthorhombic compound prototype ABC-59-ab class, which combines three different atomic species in a low-symmetry structure leading to strongly anisotropic properties, including magnetism. Excitingly, one of such materials (CrSBr) has been recently isolated down to monolayers.\cite{arXiv61}

 In turn, {\em Liu et al.} propose a layered multiferroic (MoCr$_2$S$_6$) by alloying chromium into the ferroelectric 1T phase of the MoS$_2$ matrix. First-principles calculations disclose that a spontaneous symmetry breaking, depending on the Mo atom displacement, leads to a robust ferroelectricity, which coexists with a ferromagnetic order originated from two neighboring chromium atoms. Their findings shed new light on the fundamental understanding of multiferroics and display promising applications in
spintronics and multistate data storage.\cite{liu}

Last but not least, an authoritative Tutorial contributed by {\em May et al.}\cite{may} covers the technical aspects associated with the growth of layered (anisotropic) materials {\em via} melt-based techniques, vapor transport growth, and the characterization of crystal quality with an emphasis on structural and chemical homogeneities. Important for the development of this field, details on growth and characterization of many specific compounds were provided. The tutorial's goal is to motivate more researchers to grow van der Waals crystals.

\section{Conclusions}\label{sec:iii}

In conclusion, this Special Issue showcases recent research in the rapidly-evolving area of two-dimensional materials with low-symmetry. The contributed works feature predictions of novel phases and detailed experimental discussions of growth and characterization of these phases---including the creation of van der Waals heterostructures. The variety of these published contributions is a testament to the vitality of this field.

\begin{acknowledgments}
The guest editors express their thanks to the staff and editors of the {\em Journal of Applied Physics}, as well as to the contributing authors for making the ``Beyond Graphene'' Special Topic a reality.
\end{acknowledgments}


\begin{thebibliography}{59}%
\makeatletter
\providecommand \@ifxundefined [1]{%
 \@ifx{#1\undefined}
}%
\providecommand \@ifnum [1]{%
 \ifnum #1\expandafter \@firstoftwo
 \else \expandafter \@secondoftwo
 \fi
}%
\providecommand \@ifx [1]{%
 \ifx #1\expandafter \@firstoftwo
 \else \expandafter \@secondoftwo
 \fi
}%
\providecommand \natexlab [1]{#1}%
\providecommand \enquote  [1]{``#1''}%
\providecommand \bibnamefont  [1]{#1}%
\providecommand \bibfnamefont [1]{#1}%
\providecommand \citenamefont [1]{#1}%
\providecommand \href@noop [0]{\@secondoftwo}%
\providecommand \href [0]{\begingroup \@sanitize@url \@href}%
\providecommand \@href[1]{\@@startlink{#1}\@@href}%
\providecommand \@@href[1]{\endgroup#1\@@endlink}%
\providecommand \@sanitize@url [0]{\catcode `\\12\catcode `\$12\catcode
  `\&12\catcode `\#12\catcode `\^12\catcode `\_12\catcode `\%12\relax}%
\providecommand \@@startlink[1]{}%
\providecommand \@@endlink[0]{}%
\providecommand \url  [0]{\begingroup\@sanitize@url \@url }%
\providecommand \@url [1]{\endgroup\@href {#1}{\urlprefix }}%
\providecommand \urlprefix  [0]{URL }%
\providecommand \Eprint [0]{\href }%
\providecommand \doibase [0]{http://dx.doi.org/}%
\providecommand \selectlanguage [0]{\@gobble}%
\providecommand \bibinfo  [0]{\@secondoftwo}%
\providecommand \bibfield  [0]{\@secondoftwo}%
\providecommand \translation [1]{[#1]}%
\providecommand \BibitemOpen [0]{}%
\providecommand \bibitemStop [0]{}%
\providecommand \bibitemNoStop [0]{.\EOS\space}%
\providecommand \EOS [0]{\spacefactor3000\relax}%
\providecommand \BibitemShut  [1]{\csname bibitem#1\endcsname}%
\let\auto@bib@innerbib\@empty
\bibitem [{\citenamefont {Castro~Neto}\ \emph {et~al.}(2009)\citenamefont
  {Castro~Neto}, \citenamefont {Guinea}, \citenamefont {Peres}, \citenamefont
  {Novoselov},\ and\ \citenamefont {Geim}}]{RevModPhys.81.109}%
  \BibitemOpen
  \bibfield  {author} {\bibinfo {author} {\bibfnamefont {A.~H.}\ \bibnamefont
  {Castro~Neto}}, \bibinfo {author} {\bibfnamefont {F.}~\bibnamefont {Guinea}},
  \bibinfo {author} {\bibfnamefont {N.~M.~R.}\ \bibnamefont {Peres}}, \bibinfo
  {author} {\bibfnamefont {K.~S.}\ \bibnamefont {Novoselov}}, \ and\ \bibinfo
  {author} {\bibfnamefont {A.~K.}\ \bibnamefont {Geim}},\ }\bibfield  {title}
  {\enquote {\bibinfo {title} {{{The Electronic Properties of Graphene}}},}\
  }\href {\doibase 10.1103/RevModPhys.81.109} {\bibfield  {journal} {\bibinfo
  {journal} {Rev. Mod. Phys.}\ }\textbf {\bibinfo {volume} {81}},\ \bibinfo
  {pages} {109--162} (\bibinfo {year} {2009})}\BibitemShut {NoStop}%
\bibitem [{\citenamefont {Kotov}\ \emph {et~al.}(2012)\citenamefont {Kotov},
  \citenamefont {Uchoa}, \citenamefont {Pereira}, \citenamefont {Guinea},\ and\
  \citenamefont {Castro~Neto}}]{RMP2}%
  \BibitemOpen
  \bibfield  {author} {\bibinfo {author} {\bibfnamefont {V.~N.}\ \bibnamefont
  {Kotov}}, \bibinfo {author} {\bibfnamefont {B.}~\bibnamefont {Uchoa}},
  \bibinfo {author} {\bibfnamefont {V.~M.}\ \bibnamefont {Pereira}}, \bibinfo
  {author} {\bibfnamefont {F.}~\bibnamefont {Guinea}}, \ and\ \bibinfo {author}
  {\bibfnamefont {A.~H.}\ \bibnamefont {Castro~Neto}},\ }\bibfield  {title}
  {\enquote {\bibinfo {title} {{Electron-Electron Interactions in Graphene:
  Current Status and Perspectives}},}\ }\href {\doibase
  10.1103/RevModPhys.84.1067} {\bibfield  {journal} {\bibinfo  {journal} {Rev.
  Mod. Phys.}\ }\textbf {\bibinfo {volume} {84}},\ \bibinfo {pages}
  {1067--1125} (\bibinfo {year} {2012})}\BibitemShut {NoStop}%
\bibitem [{\citenamefont {Avsar}\ \emph {et~al.}(2020)\citenamefont {Avsar},
  \citenamefont {Ochoa}, \citenamefont {Guinea}, \citenamefont {\"Ozyilmaz},
  \citenamefont {van Wees},\ and\ \citenamefont {Vera-Marun}}]{RMP3}%
  \BibitemOpen
  \bibfield  {author} {\bibinfo {author} {\bibfnamefont {A.}~\bibnamefont
  {Avsar}}, \bibinfo {author} {\bibfnamefont {H.}~\bibnamefont {Ochoa}},
  \bibinfo {author} {\bibfnamefont {F.}~\bibnamefont {Guinea}}, \bibinfo
  {author} {\bibfnamefont {B.}~\bibnamefont {\"Ozyilmaz}}, \bibinfo {author}
  {\bibfnamefont {B.~J.}\ \bibnamefont {van Wees}}, \ and\ \bibinfo {author}
  {\bibfnamefont {I.~J.}\ \bibnamefont {Vera-Marun}},\ }\bibfield  {title}
  {\enquote {\bibinfo {title} {{Colloquium: Spintronics in graphene and other
  two-dimensional materials}},}\ }\href {\doibase 10.1103/RevModPhys.92.021003}
  {\bibfield  {journal} {\bibinfo  {journal} {Rev. Mod. Phys.}\ }\textbf
  {\bibinfo {volume} {92}},\ \bibinfo {pages} {021003} (\bibinfo {year}
  {2020})}\BibitemShut {NoStop}%
\bibitem [{\citenamefont {Jin}\ \emph {et~al.}(2009)\citenamefont {Jin},
  \citenamefont {Lin}, \citenamefont {Suenaga},\ and\ \citenamefont
  {Iijima}}]{hBN}%
  \BibitemOpen
  \bibfield  {author} {\bibinfo {author} {\bibfnamefont {C.}~\bibnamefont
  {Jin}}, \bibinfo {author} {\bibfnamefont {F.}~\bibnamefont {Lin}}, \bibinfo
  {author} {\bibfnamefont {K.}~\bibnamefont {Suenaga}}, \ and\ \bibinfo
  {author} {\bibfnamefont {S.}~\bibnamefont {Iijima}},\ }\bibfield  {title}
  {\enquote {\bibinfo {title} {{{Fabrication of a Freestanding Boron Nitride
  Single Layer and Its Defect Assignments}}},}\ }\href {\doibase
  10.1103/PhysRevLett.102.195505} {\bibfield  {journal} {\bibinfo  {journal}
  {Phys. Rev. Lett.}\ }\textbf {\bibinfo {volume} {102}},\ \bibinfo {pages}
  {195505} (\bibinfo {year} {2009})}\BibitemShut {NoStop}%
\bibitem [{\citenamefont {Li}\ and\ \citenamefont {Chen}(2016)}]{hBN2}%
  \BibitemOpen
  \bibfield  {author} {\bibinfo {author} {\bibfnamefont {L.~H.}\ \bibnamefont
  {Li}}\ and\ \bibinfo {author} {\bibfnamefont {Y.}~\bibnamefont {Chen}},\
  }\bibfield  {title} {\enquote {\bibinfo {title} {{Atomically Thin Boron
  Nitride: Unique Properties and Applications}},}\ }\href {\doibase
  10.1002/adfm.201504606} {\bibfield  {journal} {\bibinfo  {journal} {Adv.
  Funct. Mater.}\ }\textbf {\bibinfo {volume} {26}},\ \bibinfo {pages}
  {2594--2608} (\bibinfo {year} {2016})}\BibitemShut {NoStop}%
\bibitem [{\citenamefont {Mak}\ \emph {et~al.}(2010)\citenamefont {Mak},
  \citenamefont {Lee}, \citenamefont {Hone}, \citenamefont {Shan},\ and\
  \citenamefont {Heinz}}]{mak1}%
  \BibitemOpen
  \bibfield  {author} {\bibinfo {author} {\bibfnamefont {K.~F.}\ \bibnamefont
  {Mak}}, \bibinfo {author} {\bibfnamefont {C.}~\bibnamefont {Lee}}, \bibinfo
  {author} {\bibfnamefont {J.}~\bibnamefont {Hone}}, \bibinfo {author}
  {\bibfnamefont {J.}~\bibnamefont {Shan}}, \ and\ \bibinfo {author}
  {\bibfnamefont {T.~F.}\ \bibnamefont {Heinz}},\ }\bibfield  {title} {\enquote
  {\bibinfo {title} {Atomically thin ${\mathrm{mos}}_{2}$: A new direct-gap
  semiconductor},}\ }\href {\doibase 10.1103/PhysRevLett.105.136805} {\bibfield
   {journal} {\bibinfo  {journal} {Phys. Rev. Lett.}\ }\textbf {\bibinfo
  {volume} {105}},\ \bibinfo {pages} {136805} (\bibinfo {year}
  {2010})}\BibitemShut {NoStop}%
\bibitem [{\citenamefont {Katsnelson}(2012)}]{Katsnelson}%
  \BibitemOpen
  \bibfield  {author} {\bibinfo {author} {\bibfnamefont {M.~I.}\ \bibnamefont
  {Katsnelson}},\ }\href@noop {} {\emph {\bibinfo {title} {{Graphene: Carbon in
  Two Dimensions}}}},\ \bibinfo {edition} {1st}\ ed.\ (\bibinfo  {publisher}
  {Cambridge U. Press},\ \bibinfo {address} {Cambdridge, UK},\ \bibinfo {year}
  {2012})\BibitemShut {NoStop}%
\bibitem [{\citenamefont {Avouris}, \citenamefont {Heinz},\ and\ \citenamefont
  {Low}(2017)}]{avouris}%
  \BibitemOpen
  \bibfield  {author} {\bibinfo {author} {\bibfnamefont {P.}~\bibnamefont
  {Avouris}}, \bibinfo {author} {\bibfnamefont {T.}~\bibnamefont {Heinz}}, \
  and\ \bibinfo {author} {\bibfnamefont {T.}~\bibnamefont {Low}},\ }\href@noop
  {} {\emph {\bibinfo {title} {{2D Materials}}}},\ \bibinfo {edition} {1st}\
  ed.\ (\bibinfo  {publisher} {Cambridge U. Press},\ \bibinfo {address}
  {Cambdridge, UK},\ \bibinfo {year} {2017})\BibitemShut {NoStop}%
\bibitem [{\citenamefont {Kolobov}\ and\ \citenamefont
  {Tominaga}(2016)}]{TMDCs}%
  \BibitemOpen
  \bibfield  {author} {\bibinfo {author} {\bibfnamefont {A.~V.}\ \bibnamefont
  {Kolobov}}\ and\ \bibinfo {author} {\bibfnamefont {J.}~\bibnamefont
  {Tominaga}},\ }\href@noop {} {\emph {\bibinfo {title} {{Two-Dimensional
  Transition-Metal Dichalcogenides}}}},\ \bibinfo {edition} {1st}\ ed.\
  (\bibinfo  {publisher} {Springer},\ \bibinfo {address} {Switzerland},\
  \bibinfo {year} {2016})\BibitemShut {NoStop}%
\bibitem [{\citenamefont {Yamasue}\ and\ \citenamefont {Cho}(2020)}]{doping}%
  \BibitemOpen
  \bibfield  {author} {\bibinfo {author} {\bibfnamefont {K.}~\bibnamefont
  {Yamasue}}\ and\ \bibinfo {author} {\bibfnamefont {Y.}~\bibnamefont {Cho}},\
  }\bibfield  {title} {\enquote {\bibinfo {title} {{Nanoscale characterization
  of unintentional doping of atomically thin layered semiconductors by scanning
  nonlinear dielectric microscopy}},}\ }\href {\doibase 10.1063/5.0016462}
  {\bibfield  {journal} {\bibinfo  {journal} {J. Appl. Phys.}\ }\textbf
  {\bibinfo {volume} {128}},\ \bibinfo {pages} {074301} (\bibinfo {year}
  {2020})}\BibitemShut {NoStop}%
\bibitem [{\citenamefont {Vozmediano}, \citenamefont {Katsnelson},\ and\
  \citenamefont {Guinea}(2010)}]{VOZMEDIANO2010109}%
  \BibitemOpen
  \bibfield  {author} {\bibinfo {author} {\bibfnamefont {M.}~\bibnamefont
  {Vozmediano}}, \bibinfo {author} {\bibfnamefont {M.}~\bibnamefont
  {Katsnelson}}, \ and\ \bibinfo {author} {\bibfnamefont {F.}~\bibnamefont
  {Guinea}},\ }\bibfield  {title} {\enquote {\bibinfo {title} {Gauge fields in
  graphene},}\ }\href {\doibase 10.1016/j.physrep.2010.07.003} {\bibfield
  {journal} {\bibinfo  {journal} {Phys. Rep.}\ }\textbf {\bibinfo {volume}
  {496}},\ \bibinfo {pages} {109--148} (\bibinfo {year} {2010})}\BibitemShut
  {NoStop}%
\bibitem [{\citenamefont {Amorim}\ \emph {et~al.}(2016)\citenamefont {Amorim},
  \citenamefont {Cortijo}, \citenamefont {{de Juan}}, \citenamefont {Grushin},
  \citenamefont {Guinea}, \citenamefont {Guti\'errez-Rubio}, \citenamefont
  {Ochoa}, \citenamefont {Parente}, \citenamefont {Roldán}, \citenamefont
  {San-Jose}, \citenamefont {Schiefele}, \citenamefont {Sturla},\ and\
  \citenamefont {Vozmediano}}]{maria}%
  \BibitemOpen
  \bibfield  {author} {\bibinfo {author} {\bibfnamefont {B.}~\bibnamefont
  {Amorim}}, \bibinfo {author} {\bibfnamefont {A.}~\bibnamefont {Cortijo}},
  \bibinfo {author} {\bibfnamefont {F.}~\bibnamefont {{de Juan}}}, \bibinfo
  {author} {\bibfnamefont {A.}~\bibnamefont {Grushin}}, \bibinfo {author}
  {\bibfnamefont {F.}~\bibnamefont {Guinea}}, \bibinfo {author} {\bibfnamefont
  {A.}~\bibnamefont {Guti\'errez-Rubio}}, \bibinfo {author} {\bibfnamefont
  {H.}~\bibnamefont {Ochoa}}, \bibinfo {author} {\bibfnamefont
  {V.}~\bibnamefont {Parente}}, \bibinfo {author} {\bibfnamefont
  {R.}~\bibnamefont {Roldán}}, \bibinfo {author} {\bibfnamefont
  {P.}~\bibnamefont {San-Jose}}, \bibinfo {author} {\bibfnamefont
  {J.}~\bibnamefont {Schiefele}}, \bibinfo {author} {\bibfnamefont
  {M.}~\bibnamefont {Sturla}}, \ and\ \bibinfo {author} {\bibfnamefont
  {M.}~\bibnamefont {Vozmediano}},\ }\bibfield  {title} {\enquote {\bibinfo
  {title} {Novel effects of strains in graphene and other two dimensional
  materials},}\ }\href {\doibase 10.1016/j.physrep.2015.12.006} {\bibfield
  {journal} {\bibinfo  {journal} {Phys. Rep.}\ }\textbf {\bibinfo {volume}
  {617}},\ \bibinfo {pages} {1--54} (\bibinfo {year} {2016})}\BibitemShut
  {NoStop}%
\bibitem [{\citenamefont {Naumis}\ \emph {et~al.}(2017)\citenamefont {Naumis},
  \citenamefont {Barraza-Lopez}, \citenamefont {Oliva-Leyva},\ and\
  \citenamefont {Terrones}}]{Review}%
  \BibitemOpen
  \bibfield  {author} {\bibinfo {author} {\bibfnamefont {G.~G.}\ \bibnamefont
  {Naumis}}, \bibinfo {author} {\bibfnamefont {S.}~\bibnamefont
  {Barraza-Lopez}}, \bibinfo {author} {\bibfnamefont {M.}~\bibnamefont
  {Oliva-Leyva}}, \ and\ \bibinfo {author} {\bibfnamefont {H.}~\bibnamefont
  {Terrones}},\ }\bibfield  {title} {\enquote {\bibinfo {title} {{{Electronic
  and Optical Properties of Strained Graphene and Other Strained 2D Materials:
  A Review}}},}\ }\href {\doibase https://doi.org/10.1088/1361-6633/aa74ef}
  {\bibfield  {journal} {\bibinfo  {journal} {Rep. Prog. Phys.}\ }\textbf
  {\bibinfo {volume} {80}},\ \bibinfo {pages} {096501} (\bibinfo {year}
  {2017})}\BibitemShut {NoStop}%
\bibitem [{\citenamefont {Huertas-Hernando}, \citenamefont {Guinea},\ and\
  \citenamefont {Brataas}(2006)}]{PhysRevB.74.155426}%
  \BibitemOpen
  \bibfield  {author} {\bibinfo {author} {\bibfnamefont {D.}~\bibnamefont
  {Huertas-Hernando}}, \bibinfo {author} {\bibfnamefont {F.}~\bibnamefont
  {Guinea}}, \ and\ \bibinfo {author} {\bibfnamefont {A.}~\bibnamefont
  {Brataas}},\ }\bibfield  {title} {\enquote {\bibinfo {title} {Spin-orbit
  coupling in curved graphene, fullerenes, nanotubes, and nanotube caps},}\
  }\href {\doibase 10.1103/PhysRevB.74.155426} {\bibfield  {journal} {\bibinfo
  {journal} {Phys. Rev. B}\ }\textbf {\bibinfo {volume} {74}},\ \bibinfo
  {pages} {155426} (\bibinfo {year} {2006})}\BibitemShut {NoStop}%
\bibitem [{\citenamefont {Duerloo}, \citenamefont {Ong},\ and\ \citenamefont
  {Reed}(2012)}]{duerloo}%
  \BibitemOpen
  \bibfield  {author} {\bibinfo {author} {\bibfnamefont {K.-A.~N.}\
  \bibnamefont {Duerloo}}, \bibinfo {author} {\bibfnamefont {M.~T.}\
  \bibnamefont {Ong}}, \ and\ \bibinfo {author} {\bibfnamefont {E.~J.}\
  \bibnamefont {Reed}},\ }\bibfield  {title} {\enquote {\bibinfo {title}
  {{Intrinsic Piezoelectricity in Two-Dimensional Materials}},}\ }\href
  {\doibase 10.1021/jz3012436} {\bibfield  {journal} {\bibinfo  {journal} {J.
  Phys. Chem. Lett.}\ }\textbf {\bibinfo {volume} {3}},\ \bibinfo {pages}
  {2871--2876} (\bibinfo {year} {2012})}\BibitemShut {NoStop}%
\bibitem [{\citenamefont {Ares}\ \emph {et~al.}(2020)\citenamefont {Ares},
  \citenamefont {Cea}, \citenamefont {Holwill}, \citenamefont {Wang},
  \citenamefont {Roldán}, \citenamefont {Guinea}, \citenamefont {Andreeva},
  \citenamefont {Fumagalli}, \citenamefont {Novoselov},\ and\ \citenamefont
  {Woods}}]{piezohBN}%
  \BibitemOpen
  \bibfield  {author} {\bibinfo {author} {\bibfnamefont {P.}~\bibnamefont
  {Ares}}, \bibinfo {author} {\bibfnamefont {T.}~\bibnamefont {Cea}}, \bibinfo
  {author} {\bibfnamefont {M.}~\bibnamefont {Holwill}}, \bibinfo {author}
  {\bibfnamefont {Y.~B.}\ \bibnamefont {Wang}}, \bibinfo {author}
  {\bibfnamefont {R.}~\bibnamefont {Roldán}}, \bibinfo {author} {\bibfnamefont
  {F.}~\bibnamefont {Guinea}}, \bibinfo {author} {\bibfnamefont {D.~V.}\
  \bibnamefont {Andreeva}}, \bibinfo {author} {\bibfnamefont {L.}~\bibnamefont
  {Fumagalli}}, \bibinfo {author} {\bibfnamefont {K.~S.}\ \bibnamefont
  {Novoselov}}, \ and\ \bibinfo {author} {\bibfnamefont {C.~R.}\ \bibnamefont
  {Woods}},\ }\bibfield  {title} {\enquote {\bibinfo {title} {{Piezoelectricity
  in Monolayer Hexagonal Boron Nitride}},}\ }\href {\doibase
  10.1002/adma.201905504} {\bibfield  {journal} {\bibinfo  {journal} {Adv.
  Mater.}\ }\textbf {\bibinfo {volume} {32}},\ \bibinfo {pages} {1905504}
  (\bibinfo {year} {2020})}\BibitemShut {NoStop}%
\bibitem [{\citenamefont {Wu}\ \emph {et~al.}(2014)\citenamefont {Wu},
  \citenamefont {Wang}, \citenamefont {Li}, \citenamefont {Zhang},
  \citenamefont {Lin}, \citenamefont {Niu}, \citenamefont {Chenet},
  \citenamefont {Zhang}, \citenamefont {Hao}, \citenamefont {Heinz},
  \citenamefont {Hone},\ and\ \citenamefont {Wang}}]{piezoMoS2}%
  \BibitemOpen
  \bibfield  {author} {\bibinfo {author} {\bibfnamefont {W.}~\bibnamefont
  {Wu}}, \bibinfo {author} {\bibfnamefont {L.}~\bibnamefont {Wang}}, \bibinfo
  {author} {\bibfnamefont {Y.}~\bibnamefont {Li}}, \bibinfo {author}
  {\bibfnamefont {F.}~\bibnamefont {Zhang}}, \bibinfo {author} {\bibfnamefont
  {L.}~\bibnamefont {Lin}}, \bibinfo {author} {\bibfnamefont {S.}~\bibnamefont
  {Niu}}, \bibinfo {author} {\bibfnamefont {D.}~\bibnamefont {Chenet}},
  \bibinfo {author} {\bibfnamefont {X.}~\bibnamefont {Zhang}}, \bibinfo
  {author} {\bibfnamefont {Y.}~\bibnamefont {Hao}}, \bibinfo {author}
  {\bibfnamefont {T.~F.}\ \bibnamefont {Heinz}}, \bibinfo {author}
  {\bibfnamefont {J.}~\bibnamefont {Hone}}, \ and\ \bibinfo {author}
  {\bibfnamefont {Z.~L.}\ \bibnamefont {Wang}},\ }\bibfield  {title} {\enquote
  {\bibinfo {title} {{Piezoelectricity of single-atomic-layer MoS$_2$ for
  energy conversion and piezotronics}},}\ }\href {\doibase 10.1038/nature13792}
  {\bibfield  {journal} {\bibinfo  {journal} {Nature}\ }\textbf {\bibinfo
  {volume} {514}},\ \bibinfo {pages} {470--474} (\bibinfo {year}
  {2014})}\BibitemShut {NoStop}%
\bibitem [{\citenamefont {Feng}\ \emph {et~al.}(2012)\citenamefont {Feng},
  \citenamefont {Qian}, \citenamefont {Huang},\ and\ \citenamefont
  {Li}}]{funnel}%
  \BibitemOpen
  \bibfield  {author} {\bibinfo {author} {\bibfnamefont {J.}~\bibnamefont
  {Feng}}, \bibinfo {author} {\bibfnamefont {X.}~\bibnamefont {Qian}}, \bibinfo
  {author} {\bibfnamefont {C.-W.}\ \bibnamefont {Huang}}, \ and\ \bibinfo
  {author} {\bibfnamefont {J.}~\bibnamefont {Li}},\ }\bibfield  {title}
  {\enquote {\bibinfo {title} {{Strain-engineered artificial atom as a
  broad-spectrum solar energy funnel}},}\ }\href {\doibase
  10.1038/nphoton.2012.285} {\bibfield  {journal} {\bibinfo  {journal} {Nat.
  Photon}\ }\textbf {\bibinfo {volume} {6}},\ \bibinfo {pages} {866--872}
  (\bibinfo {year} {2012})}\BibitemShut {NoStop}%
\bibitem [{\citenamefont {Cao}\ \emph {et~al.}(2018)\citenamefont {Cao},
  \citenamefont {Fatemi}, \citenamefont {Fang}, \citenamefont {Watanabe},
  \citenamefont {Taniguchi}, \citenamefont {Kaxiras},\ and\ \citenamefont
  {Jarillo-Herrero}}]{twisted}%
  \BibitemOpen
  \bibfield  {author} {\bibinfo {author} {\bibfnamefont {Y.}~\bibnamefont
  {Cao}}, \bibinfo {author} {\bibfnamefont {V.}~\bibnamefont {Fatemi}},
  \bibinfo {author} {\bibfnamefont {S.}~\bibnamefont {Fang}}, \bibinfo {author}
  {\bibfnamefont {K.}~\bibnamefont {Watanabe}}, \bibinfo {author}
  {\bibfnamefont {T.}~\bibnamefont {Taniguchi}}, \bibinfo {author}
  {\bibfnamefont {E.}~\bibnamefont {Kaxiras}}, \ and\ \bibinfo {author}
  {\bibfnamefont {P.}~\bibnamefont {Jarillo-Herrero}},\ }\bibfield  {title}
  {\enquote {\bibinfo {title} {{Unconventional superconductivity in magic-angle
  graphene superlattices}},}\ }\href {\doibase 10.1038/nature26160} {\bibfield
  {journal} {\bibinfo  {journal} {Nature}\ }\textbf {\bibinfo {volume} {556}},\
  \bibinfo {pages} {43--50} (\bibinfo {year} {2018})}\BibitemShut {NoStop}%
\bibitem [{\citenamefont {Haastrup}\ \emph {et~al.}(2018)\citenamefont
  {Haastrup}, \citenamefont {Strange}, \citenamefont {Pandey}, \citenamefont
  {Deilmann}, \citenamefont {Schmidt}, \citenamefont {Hinsche}, \citenamefont
  {Gjerding}, \citenamefont {Torelli}, \citenamefont {Larsen}, \citenamefont
  {Riis-Jensen}, \citenamefont {Gath}, \citenamefont {Jacobsen}, \citenamefont
  {Mortensen}, \citenamefont {Olsen},\ and\ \citenamefont {Thygesen}}]{C2DB}%
  \BibitemOpen
  \bibfield  {author} {\bibinfo {author} {\bibfnamefont {S.}~\bibnamefont
  {Haastrup}}, \bibinfo {author} {\bibfnamefont {M.}~\bibnamefont {Strange}},
  \bibinfo {author} {\bibfnamefont {M.}~\bibnamefont {Pandey}}, \bibinfo
  {author} {\bibfnamefont {T.}~\bibnamefont {Deilmann}}, \bibinfo {author}
  {\bibfnamefont {P.~S.}\ \bibnamefont {Schmidt}}, \bibinfo {author}
  {\bibfnamefont {N.~F.}\ \bibnamefont {Hinsche}}, \bibinfo {author}
  {\bibfnamefont {M.~N.}\ \bibnamefont {Gjerding}}, \bibinfo {author}
  {\bibfnamefont {D.}~\bibnamefont {Torelli}}, \bibinfo {author} {\bibfnamefont
  {P.~M.}\ \bibnamefont {Larsen}}, \bibinfo {author} {\bibfnamefont {A.~C.}\
  \bibnamefont {Riis-Jensen}}, \bibinfo {author} {\bibfnamefont
  {J.}~\bibnamefont {Gath}}, \bibinfo {author} {\bibfnamefont {K.~W.}\
  \bibnamefont {Jacobsen}}, \bibinfo {author} {\bibfnamefont {J.~J.}\
  \bibnamefont {Mortensen}}, \bibinfo {author} {\bibfnamefont {T.}~\bibnamefont
  {Olsen}}, \ and\ \bibinfo {author} {\bibfnamefont {K.~S.}\ \bibnamefont
  {Thygesen}},\ }\bibfield  {title} {\enquote {\bibinfo {title} {{The
  Computational 2D Materials Database: high-throughput modeling and discovery
  of atomically thin crystals}},}\ }\href {\doibase 10.1088/2053-1583/aacfc1}
  {\bibfield  {journal} {\bibinfo  {journal} {2D Mater.}\ }\textbf {\bibinfo
  {volume} {5}},\ \bibinfo {pages} {042002} (\bibinfo {year}
  {2018})}\BibitemShut {NoStop}%
\bibitem [{\citenamefont {Mounet}\ \emph {et~al.}(2018)\citenamefont {Mounet},
  \citenamefont {Gibertini}, \citenamefont {Schwaller}, \citenamefont {Campi},
  \citenamefont {Merkys}, \citenamefont {Marrazzo}, \citenamefont {Sohier},
  \citenamefont {Castelli}, \citenamefont {Cepellotti}, \citenamefont {Pizzi},\
  and\ \citenamefont {Marzari}}]{MarzariDatabase}%
  \BibitemOpen
  \bibfield  {author} {\bibinfo {author} {\bibfnamefont {N.}~\bibnamefont
  {Mounet}}, \bibinfo {author} {\bibfnamefont {M.}~\bibnamefont {Gibertini}},
  \bibinfo {author} {\bibfnamefont {P.}~\bibnamefont {Schwaller}}, \bibinfo
  {author} {\bibfnamefont {D.}~\bibnamefont {Campi}}, \bibinfo {author}
  {\bibfnamefont {A.}~\bibnamefont {Merkys}}, \bibinfo {author} {\bibfnamefont
  {A.}~\bibnamefont {Marrazzo}}, \bibinfo {author} {\bibfnamefont
  {T.}~\bibnamefont {Sohier}}, \bibinfo {author} {\bibfnamefont {I.~E.}\
  \bibnamefont {Castelli}}, \bibinfo {author} {\bibfnamefont {A.}~\bibnamefont
  {Cepellotti}}, \bibinfo {author} {\bibfnamefont {G.}~\bibnamefont {Pizzi}}, \
  and\ \bibinfo {author} {\bibfnamefont {N.}~\bibnamefont {Marzari}},\
  }\bibfield  {title} {\enquote {\bibinfo {title} {{Two-dimensional materials
  from high-throughput computational exfoliation of experimentally known
  compounds}},}\ }\href {\doibase 10.1038/s41565-017-0035-5} {\bibfield
  {journal} {\bibinfo  {journal} {Nat. Nanotechnol.}\ }\textbf {\bibinfo
  {volume} {13}},\ \bibinfo {pages} {246--252} (\bibinfo {year}
  {2018})}\BibitemShut {NoStop}%
\bibitem [{\citenamefont {Ashton}\ \emph {et~al.}(2017)\citenamefont {Ashton},
  \citenamefont {Paul}, \citenamefont {Sinnott},\ and\ \citenamefont
  {Hennig}}]{Hennig2017}%
  \BibitemOpen
  \bibfield  {author} {\bibinfo {author} {\bibfnamefont {M.}~\bibnamefont
  {Ashton}}, \bibinfo {author} {\bibfnamefont {J.}~\bibnamefont {Paul}},
  \bibinfo {author} {\bibfnamefont {S.~B.}\ \bibnamefont {Sinnott}}, \ and\
  \bibinfo {author} {\bibfnamefont {R.~G.}\ \bibnamefont {Hennig}},\ }\bibfield
   {title} {\enquote {\bibinfo {title} {{Topology-Scaling Identification of
  Layered Solids and Stable Exfoliated 2D Materials}},}\ }\href {\doibase
  10.1103/PhysRevLett.118.106101} {\bibfield  {journal} {\bibinfo  {journal}
  {Phys. Rev. Lett.}\ }\textbf {\bibinfo {volume} {118}},\ \bibinfo {pages}
  {106101} (\bibinfo {year} {2017})}\BibitemShut {NoStop}%
\bibitem [{\citenamefont {Choudhary}\ \emph {et~al.}(2017)\citenamefont
  {Choudhary}, \citenamefont {Kalish}, \citenamefont {Beams},\ and\
  \citenamefont {Tavazza}}]{scirep2017}%
  \BibitemOpen
  \bibfield  {author} {\bibinfo {author} {\bibfnamefont {K.}~\bibnamefont
  {Choudhary}}, \bibinfo {author} {\bibfnamefont {I.}~\bibnamefont {Kalish}},
  \bibinfo {author} {\bibfnamefont {R.}~\bibnamefont {Beams}}, \ and\ \bibinfo
  {author} {\bibfnamefont {F.}~\bibnamefont {Tavazza}},\ }\bibfield  {title}
  {\enquote {\bibinfo {title} {{High-throughput Identification and
  Characterization of Two-dimensional Materials using Density functional
  theory}},}\ }\href {\doibase 10.1038/s41598-017-05402-0} {\bibfield
  {journal} {\bibinfo  {journal} {Sci. Rep.}\ }\textbf {\bibinfo {volume}
  {7}},\ \bibinfo {pages} {5179} (\bibinfo {year} {2017})}\BibitemShut
  {NoStop}%
\bibitem [{\citenamefont {Cheon}\ \emph {et~al.}(2017)\citenamefont {Cheon},
  \citenamefont {Duerloo}, \citenamefont {Sendek}, \citenamefont {Porter},
  \citenamefont {Chen},\ and\ \citenamefont {Reed}}]{duerloo2D}%
  \BibitemOpen
  \bibfield  {author} {\bibinfo {author} {\bibfnamefont {G.}~\bibnamefont
  {Cheon}}, \bibinfo {author} {\bibfnamefont {K.-A.~N.}\ \bibnamefont
  {Duerloo}}, \bibinfo {author} {\bibfnamefont {A.~D.}\ \bibnamefont {Sendek}},
  \bibinfo {author} {\bibfnamefont {C.}~\bibnamefont {Porter}}, \bibinfo
  {author} {\bibfnamefont {Y.}~\bibnamefont {Chen}}, \ and\ \bibinfo {author}
  {\bibfnamefont {E.~J.}\ \bibnamefont {Reed}},\ }\bibfield  {title} {\enquote
  {\bibinfo {title} {{Data Mining for New Two- and One-Dimensional Weakly
  Bonded Solids and Lattice-Commensurate Heterostructures}},}\ }\href {\doibase
  10.1021/acs.nanolett.6b05229} {\bibfield  {journal} {\bibinfo  {journal}
  {Nano Lett.}\ }\textbf {\bibinfo {volume} {17}},\ \bibinfo {pages}
  {1915--1923} (\bibinfo {year} {2017})}\BibitemShut {NoStop}%
\bibitem [{\citenamefont {Zhou}\ \emph {et~al.}(2019)\citenamefont {Zhou},
  \citenamefont {Shen}, \citenamefont {Dias~Costa}, \citenamefont {Persson},
  \citenamefont {Ong}, \citenamefont {Huck}, \citenamefont {Lu}, \citenamefont
  {Ma}, \citenamefont {Chen}, \citenamefont {Tang},\ and\ \citenamefont
  {Feng}}]{2DMatPedia}%
  \BibitemOpen
  \bibfield  {author} {\bibinfo {author} {\bibfnamefont {J.}~\bibnamefont
  {Zhou}}, \bibinfo {author} {\bibfnamefont {L.}~\bibnamefont {Shen}}, \bibinfo
  {author} {\bibfnamefont {M.}~\bibnamefont {Dias~Costa}}, \bibinfo {author}
  {\bibfnamefont {K.~A.}\ \bibnamefont {Persson}}, \bibinfo {author}
  {\bibfnamefont {S.~P.}\ \bibnamefont {Ong}}, \bibinfo {author} {\bibfnamefont
  {P.}~\bibnamefont {Huck}}, \bibinfo {author} {\bibfnamefont {Y.}~\bibnamefont
  {Lu}}, \bibinfo {author} {\bibfnamefont {X.}~\bibnamefont {Ma}}, \bibinfo
  {author} {\bibfnamefont {Y.}~\bibnamefont {Chen}}, \bibinfo {author}
  {\bibfnamefont {H.}~\bibnamefont {Tang}}, \ and\ \bibinfo {author}
  {\bibfnamefont {Y.~P.}\ \bibnamefont {Feng}},\ }\bibfield  {title} {\enquote
  {\bibinfo {title} {{2DMatPedia, an open computational database of
  two-dimensional materials from top-down and bottom-up approaches}},}\ }\href
  {\doibase 10.1038/s41597-019-0097-3} {\bibfield  {journal} {\bibinfo
  {journal} {Sci. Data}\ }\textbf {\bibinfo {volume} {6}},\ \bibinfo {pages}
  {86} (\bibinfo {year} {2019})}\BibitemShut {NoStop}%
\bibitem [{\citenamefont {Kim}\ \emph {et~al.}(1997)\citenamefont {Kim},
  \citenamefont {Park}, \citenamefont {Yamaguchi}, \citenamefont {Shiino},
  \citenamefont {Kitazawa},\ and\ \citenamefont {Hasegawa}}]{tmd0}%
  \BibitemOpen
  \bibfield  {author} {\bibinfo {author} {\bibfnamefont {J.-J.}\ \bibnamefont
  {Kim}}, \bibinfo {author} {\bibfnamefont {C.}~\bibnamefont {Park}}, \bibinfo
  {author} {\bibfnamefont {W.}~\bibnamefont {Yamaguchi}}, \bibinfo {author}
  {\bibfnamefont {O.}~\bibnamefont {Shiino}}, \bibinfo {author} {\bibfnamefont
  {K.}~\bibnamefont {Kitazawa}}, \ and\ \bibinfo {author} {\bibfnamefont
  {T.}~\bibnamefont {Hasegawa}},\ }\bibfield  {title} {\enquote {\bibinfo
  {title} {{{Observation of a Phase Transition from the $T$ Phase to the $H$
  Phase Induced by a STM Tip in $1T$-TaS$_2$}}},}\ }\href {\doibase
  10.1103/PhysRevB.56.R15573} {\bibfield  {journal} {\bibinfo  {journal} {Phys.
  Rev. B}\ }\textbf {\bibinfo {volume} {56}},\ \bibinfo {pages}
  {R15573--R15576} (\bibinfo {year} {1997})}\BibitemShut {NoStop}%
\bibitem [{\citenamefont {Zhang}\ \emph {et~al.}(1996)\citenamefont {Zhang},
  \citenamefont {Liu}, \citenamefont {Huang}, \citenamefont {Kim},\ and\
  \citenamefont {Lieber}}]{tmdm1}%
  \BibitemOpen
  \bibfield  {author} {\bibinfo {author} {\bibfnamefont {J.}~\bibnamefont
  {Zhang}}, \bibinfo {author} {\bibfnamefont {J.}~\bibnamefont {Liu}}, \bibinfo
  {author} {\bibfnamefont {J.~L.}\ \bibnamefont {Huang}}, \bibinfo {author}
  {\bibfnamefont {P.}~\bibnamefont {Kim}}, \ and\ \bibinfo {author}
  {\bibfnamefont {C.~M.}\ \bibnamefont {Lieber}},\ }\bibfield  {title}
  {\enquote {\bibinfo {title} {{{Creation of Nanocrystals Through a Solid-Solid
  Phase Transition Induced by an STM Tip}}},}\ }\href {\doibase
  10.1126/science.274.5288.757} {\bibfield  {journal} {\bibinfo  {journal}
  {Science}\ }\textbf {\bibinfo {volume} {274}},\ \bibinfo {pages} {757--760}
  (\bibinfo {year} {1996})}\BibitemShut {NoStop}%
\bibitem [{\citenamefont {Chi}\ \emph {et~al.}(2018)\citenamefont {Chi},
  \citenamefont {Chen}, \citenamefont {Yen}, \citenamefont {Peng},
  \citenamefont {Zhou}, \citenamefont {Zhu}, \citenamefont {Zhang},
  \citenamefont {Liu}, \citenamefont {Lin}, \citenamefont {Chu}, \citenamefont
  {Li}, \citenamefont {Zhao}, \citenamefont {Kagayama}, \citenamefont {Ma},\
  and\ \citenamefont {Yang}}]{tp_super}%
  \BibitemOpen
  \bibfield  {author} {\bibinfo {author} {\bibfnamefont {Z.}~\bibnamefont
  {Chi}}, \bibinfo {author} {\bibfnamefont {X.}~\bibnamefont {Chen}}, \bibinfo
  {author} {\bibfnamefont {F.}~\bibnamefont {Yen}}, \bibinfo {author}
  {\bibfnamefont {F.}~\bibnamefont {Peng}}, \bibinfo {author} {\bibfnamefont
  {Y.}~\bibnamefont {Zhou}}, \bibinfo {author} {\bibfnamefont {J.}~\bibnamefont
  {Zhu}}, \bibinfo {author} {\bibfnamefont {Y.}~\bibnamefont {Zhang}}, \bibinfo
  {author} {\bibfnamefont {X.}~\bibnamefont {Liu}}, \bibinfo {author}
  {\bibfnamefont {C.}~\bibnamefont {Lin}}, \bibinfo {author} {\bibfnamefont
  {S.}~\bibnamefont {Chu}}, \bibinfo {author} {\bibfnamefont {Y.}~\bibnamefont
  {Li}}, \bibinfo {author} {\bibfnamefont {J.}~\bibnamefont {Zhao}}, \bibinfo
  {author} {\bibfnamefont {T.}~\bibnamefont {Kagayama}}, \bibinfo {author}
  {\bibfnamefont {Y.}~\bibnamefont {Ma}}, \ and\ \bibinfo {author}
  {\bibfnamefont {Z.}~\bibnamefont {Yang}},\ }\bibfield  {title} {\enquote
  {\bibinfo {title} {{Superconductivity in Pristine
  $2{H}_{a}\text{\ensuremath{-}}{\mathrm{MoS}}_{2}$ at Ultrahigh Pressure}},}\
  }\href {\doibase 10.1103/PhysRevLett.120.037002} {\bibfield  {journal}
  {\bibinfo  {journal} {Phys. Rev. Lett.}\ }\textbf {\bibinfo {volume} {120}},\
  \bibinfo {pages} {037002} (\bibinfo {year} {2018})}\BibitemShut {NoStop}%
\bibitem [{\citenamefont {Calandra}\ and\ \citenamefont
  {Mauri}(2011)}]{tp_cdw}%
  \BibitemOpen
  \bibfield  {author} {\bibinfo {author} {\bibfnamefont {M.}~\bibnamefont
  {Calandra}}\ and\ \bibinfo {author} {\bibfnamefont {F.}~\bibnamefont
  {Mauri}},\ }\bibfield  {title} {\enquote {\bibinfo {title} {{Charge-Density
  Wave and Superconducting Dome in ${\mathrm{TiSe}}_{2}$ from Electron-Phonon
  Interaction}},}\ }\href {\doibase 10.1103/PhysRevLett.106.196406} {\bibfield
  {journal} {\bibinfo  {journal} {Phys. Rev. Lett.}\ }\textbf {\bibinfo
  {volume} {106}},\ \bibinfo {pages} {196406} (\bibinfo {year}
  {2011})}\BibitemShut {NoStop}%
\bibitem [{\citenamefont {Qian}\ \emph {et~al.}(2014)\citenamefont {Qian},
  \citenamefont {Liu}, \citenamefont {Fu},\ and\ \citenamefont
  {Li}}]{qian_science_2014}%
  \BibitemOpen
  \bibfield  {author} {\bibinfo {author} {\bibfnamefont {X.}~\bibnamefont
  {Qian}}, \bibinfo {author} {\bibfnamefont {J.}~\bibnamefont {Liu}}, \bibinfo
  {author} {\bibfnamefont {L.}~\bibnamefont {Fu}}, \ and\ \bibinfo {author}
  {\bibfnamefont {J.}~\bibnamefont {Li}},\ }\bibfield  {title} {\enquote
  {\bibinfo {title} {{{Quantum spin Hall effect in two-dimensional transition
  metal dichalcogenides}}},}\ }\href {\doibase 10.1126/science.1256815}
  {\bibfield  {journal} {\bibinfo  {journal} {Science}\ }\textbf {\bibinfo
  {volume} {346}},\ \bibinfo {pages} {1344--1347} (\bibinfo {year}
  {2014})}\BibitemShut {NoStop}%
\bibitem [{\citenamefont {Fei}\ \emph {et~al.}(2017)\citenamefont {Fei},
  \citenamefont {Palomaki}, \citenamefont {Wu}, \citenamefont {Zhao},
  \citenamefont {Cai}, \citenamefont {Sun}, \citenamefont {Nguyen},
  \citenamefont {Finney}, \citenamefont {Xu},\ and\ \citenamefont
  {Cobden}}]{fei_np_2017}%
  \BibitemOpen
  \bibfield  {author} {\bibinfo {author} {\bibfnamefont {Z.}~\bibnamefont
  {Fei}}, \bibinfo {author} {\bibfnamefont {T.}~\bibnamefont {Palomaki}},
  \bibinfo {author} {\bibfnamefont {S.}~\bibnamefont {Wu}}, \bibinfo {author}
  {\bibfnamefont {W.}~\bibnamefont {Zhao}}, \bibinfo {author} {\bibfnamefont
  {X.}~\bibnamefont {Cai}}, \bibinfo {author} {\bibfnamefont {B.}~\bibnamefont
  {Sun}}, \bibinfo {author} {\bibfnamefont {P.}~\bibnamefont {Nguyen}},
  \bibinfo {author} {\bibfnamefont {J.}~\bibnamefont {Finney}}, \bibinfo
  {author} {\bibfnamefont {X.}~\bibnamefont {Xu}}, \ and\ \bibinfo {author}
  {\bibfnamefont {D.~H.}\ \bibnamefont {Cobden}},\ }\bibfield  {title}
  {\enquote {\bibinfo {title} {Edge conduction in monolayer wte2},}\ }\href
  {\doibase 10.1038/nphys4091} {\bibfield  {journal} {\bibinfo  {journal} {Nat.
  Phys.}\ }\textbf {\bibinfo {volume} {13}},\ \bibinfo {pages} {677--682}
  (\bibinfo {year} {2017})}\BibitemShut {NoStop}%
\bibitem [{\citenamefont {Tang}\ \emph {et~al.}(2017)\citenamefont {Tang},
  \citenamefont {Zhang}, \citenamefont {Wong}, \citenamefont {Pedramrazi},
  \citenamefont {Tsai}, \citenamefont {Jia}, \citenamefont {Moritz},
  \citenamefont {Claassen}, \citenamefont {Ryu}, \citenamefont {Kahn},
  \citenamefont {Jiang}, \citenamefont {Yan}, \citenamefont {Hashimoto},
  \citenamefont {Lu}, \citenamefont {Moore}, \citenamefont {Hwang},
  \citenamefont {Hwang}, \citenamefont {Hussain}, \citenamefont {Chen},
  \citenamefont {Ugeda}, \citenamefont {Liu}, \citenamefont {Xie},
  \citenamefont {Devereaux}, \citenamefont {Crommie}, \citenamefont {Mo},\ and\
  \citenamefont {Shen}}]{tang_np_2017}%
  \BibitemOpen
  \bibfield  {author} {\bibinfo {author} {\bibfnamefont {S.}~\bibnamefont
  {Tang}}, \bibinfo {author} {\bibfnamefont {C.}~\bibnamefont {Zhang}},
  \bibinfo {author} {\bibfnamefont {D.}~\bibnamefont {Wong}}, \bibinfo {author}
  {\bibfnamefont {Z.}~\bibnamefont {Pedramrazi}}, \bibinfo {author}
  {\bibfnamefont {H.-Z.}\ \bibnamefont {Tsai}}, \bibinfo {author}
  {\bibfnamefont {C.}~\bibnamefont {Jia}}, \bibinfo {author} {\bibfnamefont
  {B.}~\bibnamefont {Moritz}}, \bibinfo {author} {\bibfnamefont
  {M.}~\bibnamefont {Claassen}}, \bibinfo {author} {\bibfnamefont
  {H.}~\bibnamefont {Ryu}}, \bibinfo {author} {\bibfnamefont {S.}~\bibnamefont
  {Kahn}}, \bibinfo {author} {\bibfnamefont {J.}~\bibnamefont {Jiang}},
  \bibinfo {author} {\bibfnamefont {H.}~\bibnamefont {Yan}}, \bibinfo {author}
  {\bibfnamefont {M.}~\bibnamefont {Hashimoto}}, \bibinfo {author}
  {\bibfnamefont {D.}~\bibnamefont {Lu}}, \bibinfo {author} {\bibfnamefont
  {R.~G.}\ \bibnamefont {Moore}}, \bibinfo {author} {\bibfnamefont {C.-C.}\
  \bibnamefont {Hwang}}, \bibinfo {author} {\bibfnamefont {C.}~\bibnamefont
  {Hwang}}, \bibinfo {author} {\bibfnamefont {Z.}~\bibnamefont {Hussain}},
  \bibinfo {author} {\bibfnamefont {Y.}~\bibnamefont {Chen}}, \bibinfo {author}
  {\bibfnamefont {M.~M.}\ \bibnamefont {Ugeda}}, \bibinfo {author}
  {\bibfnamefont {Z.}~\bibnamefont {Liu}}, \bibinfo {author} {\bibfnamefont
  {X.}~\bibnamefont {Xie}}, \bibinfo {author} {\bibfnamefont {T.~P.}\
  \bibnamefont {Devereaux}}, \bibinfo {author} {\bibfnamefont {M.~F.}\
  \bibnamefont {Crommie}}, \bibinfo {author} {\bibfnamefont {S.-K.}\
  \bibnamefont {Mo}}, \ and\ \bibinfo {author} {\bibfnamefont {Z.-X.}\
  \bibnamefont {Shen}},\ }\bibfield  {title} {\enquote {\bibinfo {title}
  {Quantum spin Hall state in monolayer 1T$'$-WTe$_2$},}\ }\href {\doibase
  10.1038/nphys4174} {\bibfield  {journal} {\bibinfo  {journal} {Nat. Phys.}\
  }\textbf {\bibinfo {volume} {13}},\ \bibinfo {pages} {683--687} (\bibinfo
  {year} {2017})}\BibitemShut {NoStop}%
\bibitem [{\citenamefont {Wu}\ \emph {et~al.}(2018)\citenamefont {Wu},
  \citenamefont {Fatemi}, \citenamefont {Gibson}, \citenamefont {Watanabe},
  \citenamefont {Taniguchi}, \citenamefont {Cava},\ and\ \citenamefont
  {Jarillo-Herrero}}]{wu_science_2018}%
  \BibitemOpen
  \bibfield  {author} {\bibinfo {author} {\bibfnamefont {S.}~\bibnamefont
  {Wu}}, \bibinfo {author} {\bibfnamefont {V.}~\bibnamefont {Fatemi}}, \bibinfo
  {author} {\bibfnamefont {Q.~D.}\ \bibnamefont {Gibson}}, \bibinfo {author}
  {\bibfnamefont {K.}~\bibnamefont {Watanabe}}, \bibinfo {author}
  {\bibfnamefont {T.}~\bibnamefont {Taniguchi}}, \bibinfo {author}
  {\bibfnamefont {R.~J.}\ \bibnamefont {Cava}}, \ and\ \bibinfo {author}
  {\bibfnamefont {P.}~\bibnamefont {Jarillo-Herrero}},\ }\bibfield  {title}
  {\enquote {\bibinfo {title} {{{Observation of the quantum spin Hall effect up
  to 100 kelvin in a monolayer crystal}}},}\ }\href {\doibase
  10.1126/science.aan6003} {\bibfield  {journal} {\bibinfo  {journal}
  {Science}\ }\textbf {\bibinfo {volume} {359}},\ \bibinfo {pages} {76--79}
  (\bibinfo {year} {2018})}\BibitemShut {NoStop}%
\bibitem [{\citenamefont {Fei}\ \emph {et~al.}(2018)\citenamefont {Fei},
  \citenamefont {Zhao}, \citenamefont {Palomaki}, \citenamefont {Sun},
  \citenamefont {Miller}, \citenamefont {Zhao}, \citenamefont {Yan},
  \citenamefont {Xu},\ and\ \citenamefont {Cobden}}]{wte2}%
  \BibitemOpen
  \bibfield  {author} {\bibinfo {author} {\bibfnamefont {Z.}~\bibnamefont
  {Fei}}, \bibinfo {author} {\bibfnamefont {W.}~\bibnamefont {Zhao}}, \bibinfo
  {author} {\bibfnamefont {T.~A.}\ \bibnamefont {Palomaki}}, \bibinfo {author}
  {\bibfnamefont {B.}~\bibnamefont {Sun}}, \bibinfo {author} {\bibfnamefont
  {M.~K.}\ \bibnamefont {Miller}}, \bibinfo {author} {\bibfnamefont
  {Z.}~\bibnamefont {Zhao}}, \bibinfo {author} {\bibfnamefont {J.}~\bibnamefont
  {Yan}}, \bibinfo {author} {\bibfnamefont {X.}~\bibnamefont {Xu}}, \ and\
  \bibinfo {author} {\bibfnamefont {D.~H.}\ \bibnamefont {Cobden}},\ }\bibfield
   {title} {\enquote {\bibinfo {title} {{{Ferroelectric Switching of a
  Two-Dimensional Metal}}},}\ }\href {\doibase 10.1038/s41586-018-0336-3}
  {\bibfield  {journal} {\bibinfo  {journal} {Nature}\ }\textbf {\bibinfo
  {volume} {560}},\ \bibinfo {pages} {336--339} (\bibinfo {year}
  {2018})}\BibitemShut {NoStop}%
\bibitem [{\citenamefont {Yuan}\ \emph {et~al.}(2019)\citenamefont {Yuan},
  \citenamefont {Luo}, \citenamefont {Chan}, \citenamefont {Xiao},
  \citenamefont {Dai}, \citenamefont {Xie},\ and\ \citenamefont
  {Hao}}]{yuan_nc_2019}%
  \BibitemOpen
  \bibfield  {author} {\bibinfo {author} {\bibfnamefont {S.}~\bibnamefont
  {Yuan}}, \bibinfo {author} {\bibfnamefont {X.}~\bibnamefont {Luo}}, \bibinfo
  {author} {\bibfnamefont {H.~L.}\ \bibnamefont {Chan}}, \bibinfo {author}
  {\bibfnamefont {C.}~\bibnamefont {Xiao}}, \bibinfo {author} {\bibfnamefont
  {Y.}~\bibnamefont {Dai}}, \bibinfo {author} {\bibfnamefont {M.}~\bibnamefont
  {Xie}}, \ and\ \bibinfo {author} {\bibfnamefont {J.}~\bibnamefont {Hao}},\
  }\bibfield  {title} {\enquote {\bibinfo {title} {{{Room-temperature
  ferroelectricity in MoTe$_2$ down to the atomic monolayer limit}}},}\ }\href
  {\doibase 10.1038/s41467-019-09669-x} {\bibfield  {journal} {\bibinfo
  {journal} {Nat. Commun.}\ }\textbf {\bibinfo {volume} {10}},\ \bibinfo
  {pages} {1775} (\bibinfo {year} {2019})}\BibitemShut {NoStop}%
\bibitem [{\citenamefont {Saha}\ \emph {et~al.}(2020)\citenamefont {Saha},
  \citenamefont {Ghosh}, \citenamefont {Mazumder}, \citenamefont {Glazyrin},\
  and\ \citenamefont {Dev~Mukherjee}}]{saha}%
  \BibitemOpen
  \bibfield  {author} {\bibinfo {author} {\bibfnamefont {P.}~\bibnamefont
  {Saha}}, \bibinfo {author} {\bibfnamefont {B.}~\bibnamefont {Ghosh}},
  \bibinfo {author} {\bibfnamefont {A.}~\bibnamefont {Mazumder}}, \bibinfo
  {author} {\bibfnamefont {K.}~\bibnamefont {Glazyrin}}, \ and\ \bibinfo
  {author} {\bibfnamefont {G.}~\bibnamefont {Dev~Mukherjee}},\ }\bibfield
  {title} {\enquote {\bibinfo {title} {{Pressure induced lattice expansion and
  phonon softening in layered ReS$_2$}},}\ }\href {\doibase 10.1063/5.0014347}
  {\bibfield  {journal} {\bibinfo  {journal} {J. Appl. Phys.}\ }\textbf
  {\bibinfo {volume} {128}},\ \bibinfo {pages} {085904} (\bibinfo {year}
  {2020})}\BibitemShut {NoStop}%
\bibitem [{\citenamefont {Plumadore}\ \emph {et~al.}(2020)\citenamefont
  {Plumadore}, \citenamefont {Al~Ezzi}, \citenamefont {Adam},\ and\
  \citenamefont {Luican-Mayer}}]{plumadore}%
  \BibitemOpen
  \bibfield  {author} {\bibinfo {author} {\bibfnamefont {R.}~\bibnamefont
  {Plumadore}}, \bibinfo {author} {\bibfnamefont {M.~M.}\ \bibnamefont
  {Al~Ezzi}}, \bibinfo {author} {\bibfnamefont {S.}~\bibnamefont {Adam}}, \
  and\ \bibinfo {author} {\bibfnamefont {A.}~\bibnamefont {Luican-Mayer}},\
  }\bibfield  {title} {\enquote {\bibinfo {title} {{Moir\'e patterns in
  graphene–rhenium disulfide vertical heterostructures}},}\ }\href {\doibase
  10.1063/5.0015643} {\bibfield  {journal} {\bibinfo  {journal} {J. Appl.
  Phys.}\ }\textbf {\bibinfo {volume} {128}},\ \bibinfo {pages} {044303}
  (\bibinfo {year} {2020})}\BibitemShut {NoStop}%
\bibitem [{\citenamefont {Kipczak}\ \emph {et~al.}(2020)\citenamefont
  {Kipczak}, \citenamefont {Grzeszczyk}, \citenamefont {Olkowska-Pucko},
  \citenamefont {Babi{\'n}ski},\ and\ \citenamefont {Molas}}]{Kipczak}%
  \BibitemOpen
  \bibfield  {author} {\bibinfo {author} {\bibfnamefont {{\L}.}~\bibnamefont
  {Kipczak}}, \bibinfo {author} {\bibfnamefont {M.}~\bibnamefont {Grzeszczyk}},
  \bibinfo {author} {\bibfnamefont {K.}~\bibnamefont {Olkowska-Pucko}},
  \bibinfo {author} {\bibfnamefont {A.}~\bibnamefont {Babi{\'n}ski}}, \ and\
  \bibinfo {author} {\bibfnamefont {M.~R.}\ \bibnamefont {Molas}},\ }\bibfield
  {title} {\enquote {\bibinfo {title} {{The optical signature of few-layer
  ReSe$_2$}},}\ }\href {\doibase 10.1063/5.0015289} {\bibfield  {journal}
  {\bibinfo  {journal} {J. Appl. Phys.}\ }\textbf {\bibinfo {volume} {128}},\
  \bibinfo {pages} {044302} (\bibinfo {year} {2020})}\BibitemShut {NoStop}%
\bibitem [{\citenamefont {Liu}\ \emph {et~al.}(2014)\citenamefont {Liu},
  \citenamefont {Neal}, \citenamefont {Zhu}, \citenamefont {Luo}, \citenamefont
  {Xu}, \citenamefont {Tom\'anek},\ and\ \citenamefont {Ye}}]{ph1}%
  \BibitemOpen
  \bibfield  {author} {\bibinfo {author} {\bibfnamefont {H.}~\bibnamefont
  {Liu}}, \bibinfo {author} {\bibfnamefont {A.~T.}\ \bibnamefont {Neal}},
  \bibinfo {author} {\bibfnamefont {Z.}~\bibnamefont {Zhu}}, \bibinfo {author}
  {\bibfnamefont {Z.}~\bibnamefont {Luo}}, \bibinfo {author} {\bibfnamefont
  {X.}~\bibnamefont {Xu}}, \bibinfo {author} {\bibfnamefont {D.}~\bibnamefont
  {Tom\'anek}}, \ and\ \bibinfo {author} {\bibfnamefont {P.~D.}\ \bibnamefont
  {Ye}},\ }\bibfield  {title} {\enquote {\bibinfo {title} {{{Phosphorene: An
  Unexplored 2D Semiconductor with a High Hole Mobility}}},}\ }\href {\doibase
  10.1021/nn501226z} {\bibfield  {journal} {\bibinfo  {journal} {ACS Nano}\
  }\textbf {\bibinfo {volume} {8}},\ \bibinfo {pages} {4033--4041} (\bibinfo
  {year} {2014})}\BibitemShut {NoStop}%
\bibitem [{\citenamefont {Li}\ \emph {et~al.}(2014)\citenamefont {Li},
  \citenamefont {Yu}, \citenamefont {Ye}, \citenamefont {Ge}, \citenamefont
  {Ou}, \citenamefont {Wu}, \citenamefont {Feng}, \citenamefont {Chen},\ and\
  \citenamefont {Zhang}}]{ph2}%
  \BibitemOpen
  \bibfield  {author} {\bibinfo {author} {\bibfnamefont {L.}~\bibnamefont
  {Li}}, \bibinfo {author} {\bibfnamefont {Y.}~\bibnamefont {Yu}}, \bibinfo
  {author} {\bibfnamefont {G.~J.}\ \bibnamefont {Ye}}, \bibinfo {author}
  {\bibfnamefont {Q.}~\bibnamefont {Ge}}, \bibinfo {author} {\bibfnamefont
  {X.}~\bibnamefont {Ou}}, \bibinfo {author} {\bibfnamefont {H.}~\bibnamefont
  {Wu}}, \bibinfo {author} {\bibfnamefont {D.}~\bibnamefont {Feng}}, \bibinfo
  {author} {\bibfnamefont {X.~H.}\ \bibnamefont {Chen}}, \ and\ \bibinfo
  {author} {\bibfnamefont {Y.}~\bibnamefont {Zhang}},\ }\bibfield  {title}
  {\enquote {\bibinfo {title} {{{Black Phosphorus Field-Effect
  Transistors}}},}\ }\href {\doibase 10.1038/nnano.2014.35} {\bibfield
  {journal} {\bibinfo  {journal} {Nat. Nanotechnol.}\ }\textbf {\bibinfo
  {volume} {9}},\ \bibinfo {pages} {372--377} (\bibinfo {year}
  {2014})}\BibitemShut {NoStop}%
\bibitem [{\citenamefont {Castellanos-Gomez}\ \emph {et~al.}(2014)\citenamefont
  {Castellanos-Gomez}, \citenamefont {Vicarelli}, \citenamefont {Prada},
  \citenamefont {Island}, \citenamefont {Narasimha-Acharya}, \citenamefont
  {Blanter}, \citenamefont {Groenendijk}, \citenamefont {Buscema},
  \citenamefont {Steele}, \citenamefont {Alvarez}, \citenamefont {Zandbergen},
  \citenamefont {Palacios},\ and\ \citenamefont {van~der Zant}}]{ph3}%
  \BibitemOpen
  \bibfield  {author} {\bibinfo {author} {\bibfnamefont {A.}~\bibnamefont
  {Castellanos-Gomez}}, \bibinfo {author} {\bibfnamefont {L.}~\bibnamefont
  {Vicarelli}}, \bibinfo {author} {\bibfnamefont {E.}~\bibnamefont {Prada}},
  \bibinfo {author} {\bibfnamefont {J.~O.}\ \bibnamefont {Island}}, \bibinfo
  {author} {\bibfnamefont {K.~L.}\ \bibnamefont {Narasimha-Acharya}}, \bibinfo
  {author} {\bibfnamefont {S.~I.}\ \bibnamefont {Blanter}}, \bibinfo {author}
  {\bibfnamefont {D.~J.}\ \bibnamefont {Groenendijk}}, \bibinfo {author}
  {\bibfnamefont {M.}~\bibnamefont {Buscema}}, \bibinfo {author} {\bibfnamefont
  {G.~A.}\ \bibnamefont {Steele}}, \bibinfo {author} {\bibfnamefont {J.~V.}\
  \bibnamefont {Alvarez}}, \bibinfo {author} {\bibfnamefont {H.~W.}\
  \bibnamefont {Zandbergen}}, \bibinfo {author} {\bibfnamefont {J.~J.}\
  \bibnamefont {Palacios}}, \ and\ \bibinfo {author} {\bibfnamefont {H.~S.~J.}\
  \bibnamefont {van~der Zant}},\ }\bibfield  {title} {\enquote {\bibinfo
  {title} {{{Isolation and characterization of Few-Layer Black Phosphorus}}},}\
  }\href {\doibase 10.1088/2053-1583/1/2/025001} {\bibfield  {journal}
  {\bibinfo  {journal} {2D Mater.}\ }\textbf {\bibinfo {volume} {1}},\ \bibinfo
  {pages} {025001} (\bibinfo {year} {2014})}\BibitemShut {NoStop}%
\bibitem [{\citenamefont {Ling}\ \emph {et~al.}(2015)\citenamefont {Ling},
  \citenamefont {Wang}, \citenamefont {Huang}, \citenamefont {Xia},\ and\
  \citenamefont {Dresselhaus}}]{Ling4523}%
  \BibitemOpen
  \bibfield  {author} {\bibinfo {author} {\bibfnamefont {X.}~\bibnamefont
  {Ling}}, \bibinfo {author} {\bibfnamefont {H.}~\bibnamefont {Wang}}, \bibinfo
  {author} {\bibfnamefont {S.}~\bibnamefont {Huang}}, \bibinfo {author}
  {\bibfnamefont {F.}~\bibnamefont {Xia}}, \ and\ \bibinfo {author}
  {\bibfnamefont {M.~S.}\ \bibnamefont {Dresselhaus}},\ }\bibfield  {title}
  {\enquote {\bibinfo {title} {{The renaissance of black phosphorus}},}\ }\href
  {\doibase 10.1073/pnas.1416581112} {\bibfield  {journal} {\bibinfo  {journal}
  {Proc. Natl. Acad. Sci. (USA)}\ }\textbf {\bibinfo {volume} {112}},\ \bibinfo
  {pages} {4523--4530} (\bibinfo {year} {2015})}\BibitemShut {NoStop}%
\bibitem [{\citenamefont {{Xia}}\ \emph {et~al.}(2019)\citenamefont {{Xia}},
  \citenamefont {{Wang}}, \citenamefont {{Hwang}}, \citenamefont {{Neto}},\
  and\ \citenamefont {{Yang}}}]{2019re}%
  \BibitemOpen
  \bibfield  {author} {\bibinfo {author} {\bibfnamefont {F.}~\bibnamefont
  {{Xia}}}, \bibinfo {author} {\bibfnamefont {H.}~\bibnamefont {{Wang}}},
  \bibinfo {author} {\bibfnamefont {J.~C.~M.}\ \bibnamefont {{Hwang}}},
  \bibinfo {author} {\bibfnamefont {A.~H.~C.}\ \bibnamefont {{Neto}}}, \ and\
  \bibinfo {author} {\bibfnamefont {L.}~\bibnamefont {{Yang}}},\ }\bibfield
  {title} {\enquote {\bibinfo {title} {{Black phosphorus and its isoelectronic
  materials}},}\ }\href {\doibase 10.1038/s42254-019-0043-5} {\bibfield
  {journal} {\bibinfo  {journal} {Nature Reviews Physics}\ }\textbf {\bibinfo
  {volume} {1}},\ \bibinfo {pages} {306--317} (\bibinfo {year}
  {2019})}\BibitemShut {NoStop}%
\bibitem [{\citenamefont {Wang}\ \emph {et~al.}(2015)\citenamefont {Wang},
  \citenamefont {Jones}, \citenamefont {Seyler}, \citenamefont {Tran},
  \citenamefont {Jia}, \citenamefont {Zhao}, \citenamefont {Wang},
  \citenamefont {Yang}, \citenamefont {Xu},\ and\ \citenamefont {Xia}}]{xia2}%
  \BibitemOpen
  \bibfield  {author} {\bibinfo {author} {\bibfnamefont {X.}~\bibnamefont
  {Wang}}, \bibinfo {author} {\bibfnamefont {A.~M.}\ \bibnamefont {Jones}},
  \bibinfo {author} {\bibfnamefont {K.~L.}\ \bibnamefont {Seyler}}, \bibinfo
  {author} {\bibfnamefont {V.}~\bibnamefont {Tran}}, \bibinfo {author}
  {\bibfnamefont {Y.}~\bibnamefont {Jia}}, \bibinfo {author} {\bibfnamefont
  {H.}~\bibnamefont {Zhao}}, \bibinfo {author} {\bibfnamefont {H.}~\bibnamefont
  {Wang}}, \bibinfo {author} {\bibfnamefont {L.}~\bibnamefont {Yang}}, \bibinfo
  {author} {\bibfnamefont {X.}~\bibnamefont {Xu}}, \ and\ \bibinfo {author}
  {\bibfnamefont {F.}~\bibnamefont {Xia}},\ }\bibfield  {title} {\enquote
  {\bibinfo {title} {{Highly anisotropic and robust excitons in monolayer black
  phosphorus}},}\ }\href {\doibase 10.1038/nnano.2015.71} {\bibfield  {journal}
  {\bibinfo  {journal} {Nat. Nanotech.}\ }\textbf {\bibinfo {volume} {10}},\
  \bibinfo {pages} {517--521} (\bibinfo {year} {2015})}\BibitemShut {NoStop}%
\bibitem [{\citenamefont {Doha}\ \emph {et~al.}(2020)\citenamefont {Doha},
  \citenamefont {Santos~Batista}, \citenamefont {Rawwagah}, \citenamefont
  {Thompson}, \citenamefont {Fereidouni}, \citenamefont {Watanabe},
  \citenamefont {Taniguchi}, \citenamefont {El-Shenawee},\ and\ \citenamefont
  {Churchill}}]{DohaBPAntenna}%
  \BibitemOpen
  \bibfield  {author} {\bibinfo {author} {\bibfnamefont {M.~H.}\ \bibnamefont
  {Doha}}, \bibinfo {author} {\bibfnamefont {J.~I.}\ \bibnamefont
  {Santos~Batista}}, \bibinfo {author} {\bibfnamefont {A.~F.}\ \bibnamefont
  {Rawwagah}}, \bibinfo {author} {\bibfnamefont {J.~P.}\ \bibnamefont
  {Thompson}}, \bibinfo {author} {\bibfnamefont {A.}~\bibnamefont
  {Fereidouni}}, \bibinfo {author} {\bibfnamefont {K.}~\bibnamefont
  {Watanabe}}, \bibinfo {author} {\bibfnamefont {T.}~\bibnamefont {Taniguchi}},
  \bibinfo {author} {\bibfnamefont {M.}~\bibnamefont {El-Shenawee}}, \ and\
  \bibinfo {author} {\bibfnamefont {H.~O.~H.}\ \bibnamefont {Churchill}},\
  }\bibfield  {title} {\enquote {\bibinfo {title} {{Integration of multi-layer
  black phosphorus into photoconductive antennas for THz emission}},}\ }\href
  {\doibase 10.1063/5.0016370} {\bibfield  {journal} {\bibinfo  {journal} {J.
  Appl. Phys.}\ }\textbf {\bibinfo {volume} {128}},\ \bibinfo {pages} {063104}
  (\bibinfo {year} {2020})}\BibitemShut {NoStop}%
\bibitem [{\citenamefont {Sibari}\ \emph {et~al.}(2020)\citenamefont {Sibari},
  \citenamefont {Kerrami}, \citenamefont {Kara},\ and\ \citenamefont
  {Benaissa}}]{sibari}%
  \BibitemOpen
  \bibfield  {author} {\bibinfo {author} {\bibfnamefont {A.}~\bibnamefont
  {Sibari}}, \bibinfo {author} {\bibfnamefont {Z.}~\bibnamefont {Kerrami}},
  \bibinfo {author} {\bibfnamefont {A.}~\bibnamefont {Kara}}, \ and\ \bibinfo
  {author} {\bibfnamefont {M.}~\bibnamefont {Benaissa}},\ }\bibfield  {title}
  {\enquote {\bibinfo {title} {{Strain-engineered p-type to n-type transition
  in mono-, bi-, and tri-layer black phosphorene}},}\ }\href {\doibase
  10.1063/1.5140360} {\bibfield  {journal} {\bibinfo  {journal} {J. Appl.
  Phys.}\ }\textbf {\bibinfo {volume} {127}},\ \bibinfo {pages} {225703}
  (\bibinfo {year} {2020})}\BibitemShut {NoStop}%
\bibitem [{\citenamefont {Betancur-Ocampo}, \citenamefont {Paredes-Rocha},\
  and\ \citenamefont {Stegmann}(2020)}]{betancur}%
  \BibitemOpen
  \bibfield  {author} {\bibinfo {author} {\bibfnamefont {Y.}~\bibnamefont
  {Betancur-Ocampo}}, \bibinfo {author} {\bibfnamefont {E.}~\bibnamefont
  {Paredes-Rocha}}, \ and\ \bibinfo {author} {\bibfnamefont {T.}~\bibnamefont
  {Stegmann}},\ }\bibfield  {title} {\enquote {\bibinfo {title} {{Phosphorene
  pnp junctions as perfect electron waveguides}},}\ }\href {\doibase
  10.1063/5.0019215} {\bibfield  {journal} {\bibinfo  {journal} {J. Appl.
  Phys.}\ }\textbf {\bibinfo {volume} {128}},\ \bibinfo {pages} {114303}
  (\bibinfo {year} {2020})}\BibitemShut {NoStop}%
\bibitem [{\citenamefont {Poh}\ \emph {et~al.}(2018)\citenamefont {Poh},
  \citenamefont {Tan}, \citenamefont {Wang}, \citenamefont {Song},
  \citenamefont {Abidi}, \citenamefont {Zhao}, \citenamefont {Dan},
  \citenamefont {Chen}, \citenamefont {Luo}, \citenamefont {Pennycook},
  \citenamefont {Castro-Neto},\ and\ \citenamefont {Loh}}]{poh_nl_2018_in2se3}%
  \BibitemOpen
  \bibfield  {author} {\bibinfo {author} {\bibfnamefont {S.~M.}\ \bibnamefont
  {Poh}}, \bibinfo {author} {\bibfnamefont {S.~J.~R.}\ \bibnamefont {Tan}},
  \bibinfo {author} {\bibfnamefont {H.}~\bibnamefont {Wang}}, \bibinfo {author}
  {\bibfnamefont {P.}~\bibnamefont {Song}}, \bibinfo {author} {\bibfnamefont
  {I.~H.}\ \bibnamefont {Abidi}}, \bibinfo {author} {\bibfnamefont
  {X.}~\bibnamefont {Zhao}}, \bibinfo {author} {\bibfnamefont {J.}~\bibnamefont
  {Dan}}, \bibinfo {author} {\bibfnamefont {J.}~\bibnamefont {Chen}}, \bibinfo
  {author} {\bibfnamefont {Z.}~\bibnamefont {Luo}}, \bibinfo {author}
  {\bibfnamefont {S.~J.}\ \bibnamefont {Pennycook}}, \bibinfo {author}
  {\bibfnamefont {A.~H.}\ \bibnamefont {Castro-Neto}}, \ and\ \bibinfo {author}
  {\bibfnamefont {K.~P.}\ \bibnamefont {Loh}},\ }\bibfield  {title} {\enquote
  {\bibinfo {title} {{{Molecular-Beam Epitaxy of Two-Dimensional In$_2$Se$_3$ and Its
  Giant Electroresistance Switching in Ferroresistive Memory Junction}}},}\
  }\href {\doibase 10.1021/acs.nanolett.8b02688} {\bibfield  {journal}
  {\bibinfo  {journal} {Nano Lett.}\ }\textbf {\bibinfo {volume} {10}},\
  \bibinfo {pages} {6340--6346} (\bibinfo {year} {2018})}\BibitemShut {NoStop}%
\bibitem [{\citenamefont {Liu}\ \emph {et~al.}(2016)\citenamefont {Liu},
  \citenamefont {You}, \citenamefont {Seyler}, \citenamefont {Li},
  \citenamefont {Yu}, \citenamefont {Lin}, \citenamefont {Wang}, \citenamefont
  {Zhou}, \citenamefont {Wang}, \citenamefont {He}, \citenamefont {Pantelides},
  \citenamefont {Zhou}, \citenamefont {Sharma}, \citenamefont {Xu},
  \citenamefont {Ajayan}, \citenamefont {Wang},\ and\ \citenamefont
  {Liu}}]{cips}%
  \BibitemOpen
  \bibfield  {author} {\bibinfo {author} {\bibfnamefont {F.}~\bibnamefont
  {Liu}}, \bibinfo {author} {\bibfnamefont {L.}~\bibnamefont {You}}, \bibinfo
  {author} {\bibfnamefont {K.~L.}\ \bibnamefont {Seyler}}, \bibinfo {author}
  {\bibfnamefont {X.}~\bibnamefont {Li}}, \bibinfo {author} {\bibfnamefont
  {P.}~\bibnamefont {Yu}}, \bibinfo {author} {\bibfnamefont {J.}~\bibnamefont
  {Lin}}, \bibinfo {author} {\bibfnamefont {X.}~\bibnamefont {Wang}}, \bibinfo
  {author} {\bibfnamefont {J.}~\bibnamefont {Zhou}}, \bibinfo {author}
  {\bibfnamefont {H.}~\bibnamefont {Wang}}, \bibinfo {author} {\bibfnamefont
  {H.}~\bibnamefont {He}}, \bibinfo {author} {\bibfnamefont {S.~T.}\
  \bibnamefont {Pantelides}}, \bibinfo {author} {\bibfnamefont
  {W.}~\bibnamefont {Zhou}}, \bibinfo {author} {\bibfnamefont {P.}~\bibnamefont
  {Sharma}}, \bibinfo {author} {\bibfnamefont {X.}~\bibnamefont {Xu}}, \bibinfo
  {author} {\bibfnamefont {P.~M.}\ \bibnamefont {Ajayan}}, \bibinfo {author}
  {\bibfnamefont {J.}~\bibnamefont {Wang}}, \ and\ \bibinfo {author}
  {\bibfnamefont {Z.}~\bibnamefont {Liu}},\ }\bibfield  {title} {\enquote
  {\bibinfo {title} {{{Room-Temperature Ferroelectricity in CuInP$_2$S$_6$
  Ultrathin Flakes}}},}\ }\href {\doibase 10.1038/ncomms12357} {\bibfield
  {journal} {\bibinfo  {journal} {Nat. Commun.}\ }\textbf {\bibinfo {volume}
  {7}},\ \bibinfo {pages} {12357} (\bibinfo {year} {2016})}\BibitemShut
  {NoStop}%
\bibitem [{\citenamefont {You}\ \emph {et~al.}(2018)\citenamefont {You},
  \citenamefont {Liu}, \citenamefont {Li}, \citenamefont {Hu}, \citenamefont
  {Zhou}, \citenamefont {Chang}, \citenamefont {Zhou}, \citenamefont {Fu},
  \citenamefont {Yuan}, \citenamefont {Dong}, \citenamefont {Fan},
  \citenamefont {Gruverman}, \citenamefont {Liu},\ and\ \citenamefont
  {Wang}}]{ba2pbcl4}%
  \BibitemOpen
  \bibfield  {author} {\bibinfo {author} {\bibfnamefont {L.}~\bibnamefont
  {You}}, \bibinfo {author} {\bibfnamefont {F.}~\bibnamefont {Liu}}, \bibinfo
  {author} {\bibfnamefont {H.}~\bibnamefont {Li}}, \bibinfo {author}
  {\bibfnamefont {Y.}~\bibnamefont {Hu}}, \bibinfo {author} {\bibfnamefont
  {S.}~\bibnamefont {Zhou}}, \bibinfo {author} {\bibfnamefont {L.}~\bibnamefont
  {Chang}}, \bibinfo {author} {\bibfnamefont {Y.}~\bibnamefont {Zhou}},
  \bibinfo {author} {\bibfnamefont {Q.}~\bibnamefont {Fu}}, \bibinfo {author}
  {\bibfnamefont {G.}~\bibnamefont {Yuan}}, \bibinfo {author} {\bibfnamefont
  {S.}~\bibnamefont {Dong}}, \bibinfo {author} {\bibfnamefont {H.~J.}\
  \bibnamefont {Fan}}, \bibinfo {author} {\bibfnamefont {A.}~\bibnamefont
  {Gruverman}}, \bibinfo {author} {\bibfnamefont {Z.}~\bibnamefont {Liu}}, \
  and\ \bibinfo {author} {\bibfnamefont {J.}~\bibnamefont {Wang}},\ }\bibfield
  {title} {\enquote {\bibinfo {title} {In-plane ferroelectricity in thin flakes
  of van der waals hybrid perovskite},}\ }\href {\doibase
  10.1002/adma.201803249} {\bibfield  {journal} {\bibinfo  {journal} {Adv.
  Mater.}\ }\textbf {\bibinfo {volume} {30}},\ \bibinfo {pages} {1803249}
  (\bibinfo {year} {2018})}\BibitemShut {NoStop}%
\bibitem [{\citenamefont {Barraza-Lopez}\ \emph {et~al.}(2020)\citenamefont
  {Barraza-Lopez}, \citenamefont {Fregoso}, \citenamefont {Villanova},
  \citenamefont {Parkin},\ and\ \citenamefont {Chang}}]{colloquium}%
  \BibitemOpen
  \bibfield  {author} {\bibinfo {author} {\bibfnamefont {S.}~\bibnamefont
  {Barraza-Lopez}}, \bibinfo {author} {\bibfnamefont {B.~M.}\ \bibnamefont
  {Fregoso}}, \bibinfo {author} {\bibfnamefont {W.~J.}\ \bibnamefont
  {Villanova}}, \bibinfo {author} {\bibfnamefont {S.~S.~P.}\ \bibnamefont
  {Parkin}}, \ and\ \bibinfo {author} {\bibfnamefont {K.}~\bibnamefont
  {Chang}},\ }\bibfield  {title} {\enquote {\bibinfo {title} {{Colloquium:
  physical behavior of group-IV monochalcogenide monolayers}},}\ }\href
  {https://arxiv.org/abs/2009.04341} {\bibfield  {journal} {\bibinfo  {journal}
  {Rev. Mod. Phys. (accepted); arXiv:2009.04341}\ } (\bibinfo {year}
  {2020})}\BibitemShut {NoStop}%
\bibitem [{\citenamefont {Chang}\ and\ \citenamefont {Parkin}(2020)}]{chang}%
  \BibitemOpen
  \bibfield  {author} {\bibinfo {author} {\bibfnamefont {K.}~\bibnamefont
  {Chang}}\ and\ \bibinfo {author} {\bibfnamefont {S.~S.~P.}\ \bibnamefont
  {Parkin}},\ }\bibfield  {title} {\enquote {\bibinfo {title} {{Experimental
  formation of monolayer group-IV monochalcogenides}},}\ }\href {\doibase
  10.1063/5.0012300} {\bibfield  {journal} {\bibinfo  {journal} {J. Appl.
  Phys.}\ }\textbf {\bibinfo {volume} {127}},\ \bibinfo {pages} {220902}
  (\bibinfo {year} {2020})}\BibitemShut {NoStop}%
\bibitem [{\citenamefont {Du}, \citenamefont {Pendergrast},\ and\ \citenamefont
  {Barraza-Lopez}(2020)}]{doping2}%
  \BibitemOpen
  \bibfield  {author} {\bibinfo {author} {\bibfnamefont {A.}~\bibnamefont
  {Du}}, \bibinfo {author} {\bibfnamefont {Z.}~\bibnamefont {Pendergrast}}, \
  and\ \bibinfo {author} {\bibfnamefont {S.}~\bibnamefont {Barraza-Lopez}},\
  }\bibfield  {title} {\enquote {\bibinfo {title} {Tuning energy barriers by
  doping 2D group-IV monochalcogenides},}\ }\href {\doibase 10.1063/5.0008502}
  {\bibfield  {journal} {\bibinfo  {journal} {J. Appl. Phys.}\ }\textbf
  {\bibinfo {volume} {127}},\ \bibinfo {pages} {234103} (\bibinfo {year}
  {2020})}\BibitemShut {NoStop}%
\bibitem [{\citenamefont {Seixas}(2020)}]{seixas}%
  \BibitemOpen
  \bibfield  {author} {\bibinfo {author} {\bibfnamefont {L.}~\bibnamefont
  {Seixas}},\ }\bibfield  {title} {\enquote {\bibinfo {title} {{Janus
  two-dimensional materials based on group IV monochalcogenides}},}\ }\href
  {\doibase 10.1063/5.0012427} {\bibfield  {journal} {\bibinfo  {journal} {J.
  Appl. Phys.}\ }\textbf {\bibinfo {volume} {128}},\ \bibinfo {pages} {045115}
  (\bibinfo {year} {2020})}\BibitemShut {NoStop}%
\bibitem [{\citenamefont {Sandoval-Santana}\ \emph {et~al.}(2020)\citenamefont
  {Sandoval-Santana}, \citenamefont {Ibarra-Sierra}, \citenamefont {Kunold},\
  and\ \citenamefont {Naumis}}]{gerardo}%
  \BibitemOpen
  \bibfield  {author} {\bibinfo {author} {\bibfnamefont {J.~C.}\ \bibnamefont
  {Sandoval-Santana}}, \bibinfo {author} {\bibfnamefont {V.~G.}\ \bibnamefont
  {Ibarra-Sierra}}, \bibinfo {author} {\bibfnamefont {A.}~\bibnamefont
  {Kunold}}, \ and\ \bibinfo {author} {\bibfnamefont {G.~G.}\ \bibnamefont
  {Naumis}},\ }\bibfield  {title} {\enquote {\bibinfo {title} {{Floquet
  spectrum for anisotropic and tilted Dirac materials under linearly polarized
  light at all field intensities}},}\ }\href {\doibase 10.1063/5.0007576}
  {\bibfield  {journal} {\bibinfo  {journal} {J. Appl. Phys.}\ }\textbf
  {\bibinfo {volume} {127}},\ \bibinfo {pages} {234301} (\bibinfo {year}
  {2020})}\BibitemShut {NoStop}%
\bibitem [{\citenamefont {Vannucci}\ \emph {et~al.}(2020)\citenamefont
  {Vannucci}, \citenamefont {Petralanda}, \citenamefont {Rasmussen},
  \citenamefont {Olsen},\ and\ \citenamefont {Thygesen}}]{1-Vannucci}%
  \BibitemOpen
  \bibfield  {author} {\bibinfo {author} {\bibfnamefont {L.}~\bibnamefont
  {Vannucci}}, \bibinfo {author} {\bibfnamefont {U.}~\bibnamefont
  {Petralanda}}, \bibinfo {author} {\bibfnamefont {A.}~\bibnamefont
  {Rasmussen}}, \bibinfo {author} {\bibfnamefont {T.}~\bibnamefont {Olsen}}, \
  and\ \bibinfo {author} {\bibfnamefont {K.~S.}\ \bibnamefont {Thygesen}},\
  }\bibfield  {title} {\enquote {\bibinfo {title} {Anisotropic properties of
  monolayer 2d materials: An overview from the C2DB database},}\ }\href
  {\doibase 10.1063/5.0021237} {\bibfield  {journal} {\bibinfo  {journal} {J.
  Appl. Phys.}\ }\textbf {\bibinfo {volume} {128}},\ \bibinfo {pages} {105101}
  (\bibinfo {year} {2020})}\BibitemShut {NoStop}%
\bibitem [{\citenamefont {Lee}\ \emph {et~al.}(2020)\citenamefont {Lee},
  \citenamefont {Dismukes}, \citenamefont {Telford}, \citenamefont {Wiscons},
  \citenamefont {Xu}, \citenamefont {Nuckolls}, \citenamefont {Dean},
  \citenamefont {Roy},\ and\ \citenamefont {Zhu}}]{arXiv61}%
  \BibitemOpen
  \bibfield  {author} {\bibinfo {author} {\bibfnamefont {K.}~\bibnamefont
  {Lee}}, \bibinfo {author} {\bibfnamefont {A.~H.}\ \bibnamefont {Dismukes}},
  \bibinfo {author} {\bibfnamefont {E.~J.}\ \bibnamefont {Telford}}, \bibinfo
  {author} {\bibfnamefont {R.~A.}\ \bibnamefont {Wiscons}}, \bibinfo {author}
  {\bibfnamefont {X.}~\bibnamefont {Xu}}, \bibinfo {author} {\bibfnamefont
  {C.}~\bibnamefont {Nuckolls}}, \bibinfo {author} {\bibfnamefont {C.~R.}\
  \bibnamefont {Dean}}, \bibinfo {author} {\bibfnamefont {X.}~\bibnamefont
  {Roy}}, \ and\ \bibinfo {author} {\bibfnamefont {X.}~\bibnamefont {Zhu}},\
  }\bibfield  {title} {\enquote {\bibinfo {title} {{Magnetic Order and Symmetry
  in the 2D Semiconductor CrSBr}},}\ }\href {http://arxiv.org/abs/2007.10715}
  {\bibfield  {journal} {\bibinfo  {journal} {arXiv:2007.10715}\ } (\bibinfo
  {year} {2020})}\BibitemShut {NoStop}%
\bibitem [{\citenamefont {Liu}, \citenamefont {Jin},\ and\ \citenamefont
  {Liu}(2020)}]{liu}%
  \BibitemOpen
  \bibfield  {author} {\bibinfo {author} {\bibfnamefont {L.-Z.}\ \bibnamefont
  {Liu}}, \bibinfo {author} {\bibfnamefont {K.-H.}\ \bibnamefont {Jin}}, \ and\
  \bibinfo {author} {\bibfnamefont {F.}~\bibnamefont {Liu}},\ }\bibfield
  {title} {\enquote {\bibinfo {title} {{Prediction of room-temperature
  multiferroicity in strained MoCr$_2$S$_6$ monolayer}},}\ }\href {\doibase
  10.1063/1.5144535} {\bibfield  {journal} {\bibinfo  {journal} {J. Appl.
  Phys.}\ }\textbf {\bibinfo {volume} {127}},\ \bibinfo {pages} {155302}
  (\bibinfo {year} {2020})}\BibitemShut {NoStop}%
\bibitem [{\citenamefont {May}, \citenamefont {Yan},\ and\ \citenamefont
  {McGuire}(2020)}]{may}%
  \BibitemOpen
  \bibfield  {author} {\bibinfo {author} {\bibfnamefont {A.~F.}\ \bibnamefont
  {May}}, \bibinfo {author} {\bibfnamefont {J.}~\bibnamefont {Yan}}, \ and\
  \bibinfo {author} {\bibfnamefont {M.~A.}\ \bibnamefont {McGuire}},\
  }\bibfield  {title} {\enquote {\bibinfo {title} {{A practical guide for
  crystal growth of van der Waals layered materials}},}\ }\href {\doibase
  10.1063/5.0015971} {\bibfield  {journal} {\bibinfo  {journal} {J. Appl.
  Phys.}\ }\textbf {\bibinfo {volume} {128}},\ \bibinfo {pages} {051101}
  (\bibinfo {year} {2020})}\BibitemShut {NoStop}%
\end{thebibliography}

%

\end{document}